\documentclass{emulateapj}
\usepackage{apjfonts}

\newcommand{\bmbx}{BM/BX}
\newcommand{\bx}{BX}
\newcommand{\etal}{et al.\,}

\slugcomment{Accepted for publication in the Astrophysical Journal}

\shorttitle{Integral field spectroscopy of $z \sim 2$ galaxies}
\shortauthors{F\"orster Schreiber \etal}

\begin{document}

\title{
SINFONI Integral Field Spectroscopy of $z \sim 2$ UV-selected Galaxies:
Rotation Curves and Dynamical Evolution\,\altaffilmark{1}}

\author{N. M. F\"orster Schreiber\altaffilmark{2},
        R. Genzel\altaffilmark{2,3},
        M. D. Lehnert\altaffilmark{2},
	N. Bouch\'e\altaffilmark{2},
        A. Verma\altaffilmark{2},
        D. K. Erb\altaffilmark{4,5}
        A. E. Shapley\altaffilmark{6,7},
        C. C. Steidel\altaffilmark{4},
        R. Davies\altaffilmark{2},
        D. Lutz\altaffilmark{2},
        N. Nesvadba\altaffilmark{2},
        L. J. Tacconi\altaffilmark{2},        
        F. Eisenhauer\altaffilmark{2},
        R. Abuter\altaffilmark{2},
        A. Gilbert\altaffilmark{2},
        S. Gillessen\altaffilmark{2},
        A. Sternberg\altaffilmark{8}}

\altaffiltext{1}{Based on observations obtained at the Very Large Telescope
                 (VLT) of the European Southern Observatory, Paranal, Chile
                 (ESO Programme IDs
                  073.B-9018, 074.A-9011, 075.A-0466, and 076.A-0527).}
\altaffiltext{2}{Max-Planck-Institut f\"ur extraterrestrische Physik,
                 Giessenbachstrasse, D-85748 Garching, Germany}
\altaffiltext{3}{Department of Physics, University of California,
                 Berkeley, CA 94720}
\altaffiltext{4}{California Institute of Technology, MS 105-24,
                 Pasadena, CA 91125}
\altaffiltext{5}{Present address: Harvard-Smithsonian Center for Astrophysics,
                 Cambridge, Massachusetts, MA 02138}
\altaffiltext{6}{Department of Astronomy, University of California,
                 Berkeley, CA 94720}
\altaffiltext{7}{Present address: Department of Astrophysical Sciences,
                 Princeton University, Princeton, NJ 08544-1001}
\altaffiltext{8}{School of Physics and Astronomy,
                 Tel Aviv University, Tel Aviv 69978, Israel}

\begin{abstract}

We present $\sim 0\farcs 5$ resolution near-infrared integral field
spectroscopy of the H$\alpha$ line emission of 14 $z \sim 2$ UV-selected
\bmbx\ galaxies, obtained with SINFONI at the ESO Very Large Telescope.
The average H$\alpha$ half-light radius is
$r_{1/2} \approx 4\,h_{70}^{-1}~{\rm kpc}$, and line emission is
detected over $\rm \ga 20\,h_{70}^{-1}~kpc$ in several sources.
In nine galaxies, we detect spatially-resolved velocity gradients,
from 40 to $\rm 410~km\,s^{-1}$ over $\sim 10\,h_{70}^{-1}~{\rm kpc}$.
The observed kinematics of the larger systems are generally consistent
with orbital motions.  Four galaxies are well described by rotating disks
with clumpy morphologies, and we extracted rotation curves out to radii
$\ga 10\,h_{70}^{-1}~{\rm kpc}$.  One or two galaxies exhibit signatures
more consistent with mergers.  Analyzing all 14 galaxies of the sample
in the framework of rotating disks, we infer mean inclination- and
beam-corrected maximum circular velocities of
$v_{\rm c} \sim 180 \pm 90~{\rm km\,s^{-1}}$ and dynamical masses from 
$\sim 0.5$ to $25 \times 10^{10}\,h_{70}^{-1}~{\rm M_{\odot}}$ within
$r_{1/2}$.  On average, the dynamical masses are consistent with estimates of
photometric stellar masses assuming a Chabrier or Kroupa initial mass function
(IMF) but too small for a $\rm 0.1 - 100~{\rm M_{\odot}}$ Salpeter IMF.
The specific angular momenta of our \bmbx\ galaxies are similar to those of
local late-type galaxies.  In the single disk framework, the specific angular
momenta of the baryons are comparable to those of their dark matter halos.
Extrapolating from the average $v_{\rm c}$ at $10\,h_{70}^{-1}~{\rm kpc}$,
the virial mass of the typical halo of a galaxy in our sample is
$10^{11.7 \pm 0.5}\,h_{70}^{-1}~{\rm M_{\odot}}$, in good agreement with
previous estimates from the clustering and number density of the \bmbx\
population.  Kinematic modeling of the three best cases implies a ratio of
$v_{\rm c}$ to local velocity dispersion $v_{\rm c}/\sigma \sim 2 - 4$ and
accordingly a large geometric thickness.  We argue that this suggests a mass
accretion (alternatively, gas exhaustion) timescale of $\rm \sim 500~Myr$.
We also argue that if our \bmbx\ galaxies were initially gas rich, their
clumpy disks will subsequently lose their angular momentum and form compact
bulges on a timescale of $\rm \sim 1~Gyr$.
In support of this scenario, the object with brightest $K$-band magnitude
and among the reddest in $J - K$ colour in our sample (suggesting it is
the most evolved) exhibits a maximum in [\ion{N}{2}]/H$\alpha$ ratio and
a minimum in H$\alpha$ equivalent width at the geometric/kinematic center,
which we interpret as indicative of an inside-out metallicity and age gradient.

\end{abstract}


\section{INTRODUCTION}    \label{Intro}

Over the last decade, deep surveys in the optical, infrared, and
submillimeter wavebands have revealed a large diversity among high redshift
galaxy populations.  This reflects in part a diversity in properties such as
mass, stellar populations, and gas and dust content.  High redshift samples
have become sufficiently complete to enable fairly robust estimates of the
cosmic evolution of the stellar mass and luminosity density, star formation
history, and nuclear activity.  About half the stellar mass density in
the local universe appears to have been assembled by $z \sim 1$
\citep[e.g.,][]{Dic03, Rud03, Fon03}, after the major peak of star formation
and QSO activity at $z \sim 2 - 2.5$ \citep[e.g.,][]{Fan01, Cha05}.

It is, however, less clear exactly how galaxies were assembled.
In particular, when and at what rate did galaxies of different
masses form?  What is the connection between bulge and disk formation? 
Progress in solving these questions is hampered by our incomplete
knowledge of various key physical and observational aspects.
For instance, we still have a poor understanding of the competing processes
of cooling, angular momentum exchange and loss, and feedback from star
formation and active galactic nuclei (AGN), which drive galaxy evolution.
The links between the various high redshift populations, their origin,
and their subsequent fates still have to be established.  In the framework
of semi-analytical models of structure formation, the description of the
formation and evolution of galaxies remains uncertain because of the lack
of observations constraining the parametrization of the complex physical
processes involved.  The most important ingredients include angular momentum,
metallicity, stellar initial mass function (IMF), the spatial distribution
of the gas and stars, feedback, and environmental effects such as merger
frequency and harassment.  A better knowledge of the spatially-resolved
kinematics, morphologies, stellar populations, metallicities, and ionization
state of the line-emitting gas within galaxies will greatly aid in addressing
these issues.

In this context, we have begun a substantial program to investigate
representative samples of galaxy populations at $z \sim 1 - 4$ from
integral field spectroscopy at near-infrared (NIR) wavelengths with
SINFONI at the Very Large Telescope (VLT) of the European Southern
Observatory (ESO): the high redshift galaxy Spectroscopic Imaging
survey in the Near-infrared with SINFONI (SINS).
For $z = 1 - 4$, NIR observations probe the rest-frame optical emission,
which contains numerous diagnostic spectral features that have been well
studied and calibrated from large samples at low redshift.  SINFONI enables
the full two-dimensional mapping of the spectral features, in seeing-limited
or adaptive-optics mode.  To obtain a panoramic view of the characteristic
spatially-resolved properties, we target samples of galaxies selected by a
variety of well-defined criteria.  One of the class we have observed is the
rest-frame UV-selected ``\bmbx'' galaxies, which are efficiently selected
in optical surveys from colour criteria analogous to those of the Lyman-break
technique but targeting the lower redshift range $z \sim 1.5 - 2.5$
\citep{Ade04, Ste04}.  In this paper, we present the results of a sample
of 14 \bmbx\ objects, focussing on the morphologies and kinematics traced
by the H$\alpha$ line emission.

In \S~\ref{Sect-data}, we describe the observations and data reduction of
our \bmbx\ galaxy sample.  In \S~\ref{Sect-res}, we present the results of
the H$\alpha$ morphologies and kinematics for the entire sample while in
\S~\ref{Sect-case}, we present detailed case studies of three of the objects,
which have the best quality two-dimensional information.  We discuss our
results in \S~\ref{Sect-disc} and summarize the paper in \S~\ref{Sect-conclu}.
Throughout, we assume a $\Lambda$-dominated cosmology
with $H_{0} = 70\,h_{70}~{\rm km\,s^{-1}\,Mpc^{-1}}$,
$\Omega_{\rm m} = 0.3$, and $\Omega_{\Lambda} = 0.7$,
unless explicitely stated otherwise.  For this cosmology,
1\arcsec\ corresponds to $\rm \approx 8.2~kpc$ at $z = 2.2$.
Magnitudes are given in the Vega-based photometric system.

\section{OBSERVATIONS AND DATA REDUCTION}    \label{Sect-data}

We selected our \bmbx\ targets from the samples observed with long-slit
NIR spectroscopy of \citet{Erb03}, \citet{Sha04}, \citet{Ste04}, and
\citet{Erb06b}.  Our criteria were a combination of target visibility,
night sky line avoidance for H$\alpha$, and H$\alpha$ line flux, with
emphasis on objects with spatially extended emission, spatially resolved
velocity gradients, or high velocity dispersions based on the existing
long-slit spectroscopy.  To date, we have observed 14 objects: one BM source
at $z = 1.41$ and 13 \bx\ sources in the range $z = 2.0 - 2.5$, with a mean
and median redshift of 2.24.  Our sample includes one clearly identified pair,
$\rm Q2346-BX404/405$, with projected angular separation of 3\farcs 6
(corresponding to 30~kpc at the redshift of the system).
The optical and NIR photometric properties of our targets are presented by
\citet{Erb06b} and are very similar to those of their $\sim 100$ \bmbx\ objects
with long-slit NIR spectroscopy, spanning nearly the full range observed 
for the general \bmbx\ population in terms of $\mathcal{R}$ and $K_{\rm s}$
magnitudes as well as $\mathcal{R} - K_{\rm s}$ and $J - K_{\rm s}$ colours
\citep[see, e.g.,][]{Ste04, Red05}.  For comparison with other surveys,
three of the 11 targets of our sample with available $K_{\rm s}$-band
photometry have $K_{\rm s} < 20~{\rm mag}$, and two of the eight with
additional $J$-band photometry have $J - K_{\rm s} > 2.3~{\rm mag}$.
We refer the reader to \citet{Erb06b} for a more detailed discussion.
Table~\ref{tab-obs} lists the sample (along with the redshifts derived
from H$\alpha$ in \S~\ref{Sect-res}) and summarizes the observations.

We collected the data with SINFONI \citep{Eis03a, Bon04, Gil05} mounted
at the Cassegrain focus of the VLT UT4 telescope during several observing
campaigns between 2004 July and 2005 October, as part of commissioning and
guaranteed time observations.  SINFONI consists of SPIFFI \citep{Eis03b},
a NIR cryogenic integral field spectrometer equipped with a $\rm 2K^{2}$
HAWAII detector, and a curvature-sensor adaptive optics (AO) module
\citep[MACAO;][]{Bon03}.  We used the $H$- or $K$-band gratings to map
the H$\alpha$ line emission, depending on the redshift of the galaxies.
For $\rm Q2343-BX610$, we complemented $K$-band data with observations
in the $J$ and $H$ bands to determine the abundance indicator ``$R_{23}$''
from the [\ion{O}{2}]\,$\lambda\,3727$,
[\ion{O}{3}]\,$\lambda\lambda\,4959,5007$, and H$\beta$ emission lines.
We carried out most of the observations in seeing-limited mode with the
$\rm 0\farcs 125~pixel^{-1}$ scale, for which the total field of view
(FOV) is $8^{\prime\prime } \times 8^{\prime\prime}$.
For two sources, the presence of a sufficiently bright nearby star
enabled us to obtain AO-assisted observations.  For $\rm Q1623-BX502$
(guide star 8\arcsec\ away), we used the $\rm 0\farcs 05~pixel^{-1}$
scale with total FOV of $3\farcs 2 \times 3\farcs 2$.  For $\rm Q1623-BX663$,
we employed the $\rm 0\farcs 125~pixel^{-1}$ scale and locked on the
guide star (13\arcsec\ away) for one-third of the total integration time.
In these configurations, SINFONI delivers a nominal FWHM spectral resolution
of $R \sim 4500$ in $K$ ($\rm \lambda = 1.95 - 2.45~\mu m$),
$\sim 3000$ in $H$ ($\rm \lambda = 1.45 - 1.85~\mu m$), and
$\sim 1900$ in $J$ ($\rm \lambda = 1.10 - 1.40~\mu m$).

We took the data in series of ``AB'' cycles, with typical nod throws of about
half the SINFONI FOV so as to image the source in all frames, and jitter box
widths of about one-tenth the FOV to minimize the number of redundant positions
on the detector array.  The individual exposure times varied from 300\,s to
900\,s depending mainly on the variability and intensity of the background
and night sky line emission, in order to optimize the background subtraction
and remain in the background-limited regime over the wavelength range of
interest around the H$\alpha$ line.  The total on-source integration times
ranged from 1\,hr to 8\,hr, driven by the surface brightness of the sources
and our aim to map the kinematics out to large radii for those with the most
extended H$\alpha$ emission and the largest velocity gradients.
We took exposures of stars (including the acquisition stars used for blind
offsetting to the galaxies as well as late-B, A0V, and G1V to G3V telluric
standards) interleaved with the target observations to monitor the seeing
and the atmospheric transmission.

The data reduction included the following steps.  We performed wavelength
calibration of the individual \bmbx\ exposures using arc lamp frames, which
we verified and corrected, if necessary, using night sky lines present in
the data.  We subtracted the \bmbx\ frames pairwise (without spatial shifting)
to remove the night sky line and background emission.  For some galaxies,
we applied a more sophisticated sky subtraction procedure that involves
scaling each transition group of telluric OH lines separately in order to
minimize residuals affecting the emission lines of interest of the source.
We flat-fielded the \bmbx\ data with exposures of a halogen calibration lamp.
We identified bad pixels based on dark and flat-field frames, and replaced
them using interpolation.
We then reconstructed (including a distortion correction) and registered
the three-dimensional data cubes, and averaged them together to produce
the final reduced cube for each source.
The data of the offset stars and telluric standards were reduced in a
similar way.  Based on the reference stars data, the seeing was typically
$\approx 0\farcs 5$ (FWHM) in the NIR throughout the observations.
For $\rm Q1623-BX502$, the use of AO at the $\rm 0\farcs 05~pixel^{-1}$
scale improved the effective resolution to a FWHM of 0\farcs 3.  From the
night sky lines, we determined that the effective spectral resolution of the
reduced \bmbx\ data cubes corresponds to $\rm \approx 80~km\,s^{-1}$ (FWHM).

\section{RESULTS}    \label{Sect-res}

We extracted maps of the velocity-integrated line flux, velocity field,
and FWHM velocity width (hereafter $\Delta v$) of H$\alpha$ from the reduced
data cubes of the \bmbx\ objects.  We also extracted position-velocity (p-v)
diagrams in a $\approx 0\farcs 5$-wide synthetic slit oriented along the
kinematic major axis, identified from the SINFONI data for each object.
To increase the signal-to-noise (S/N) ratio, we smoothed some of the p-v
diagrams with a two-dimensional Gaussian of $\rm FWHM = 2~pixels$, or
0\farcs 25 spatially and $\rm \approx 70~km\,s^{-1}$ spectrally.  For the
velocity and velocity width maps, we smoothed the reduced cubes spatially
with a Gaussian of $\rm FWHM = 3~pixels$, resulting in a typical effective
spatial resolution of $\approx 0\farcs 6$ (0\farcs 34 for $\rm Q1623-BX502$
observed in AO mode with the $\rm 0\farcs 05~pixel^{-1}$ scale).
We determined the total line flux, the systemic redshift, and the
total $\Delta v$ from H$\alpha$ by fitting a Gaussian profile to the 
spatially-integrated spectrum of each source.  The velocity fields and
velocity FWHM maps were extracted from the smoothed data cubes by applying
the same Gaussian fitting procedure to each spatial pixel as to the integrated
spectrum
\footnote{Throughout the paper, ``velocity'' and ``velocity width''
 thus refer to the center and FWHM parameters of the Gaussians that
 best fit the observed line profiles.}.
Except for $\rm Q2343-BX610$ (which has a $K$-band magnitude of 19.2~mag),
the continuum emission of the \bmbx\ objects in the sample is sufficiently
weak so that no continuum subtraction was necessary in making the linemaps.
We measured $v_{\rm r}$, defined as half the maximum observed velocity
gradient from Gaussian fits to spectra taken in apertures on opposite
sides of each source along the kinematic major axis.  We also estimated
the intrinsic half-light radius $r_{1/2}$ along the geometric major axis
from Gaussian fitting to the H$\alpha$ linemaps.

Table~\ref{tab-res} lists the redshift, $r_{1/2}$, $\Delta v$, and $v_{r}$
measurements.  The half-light radii and FWHM line widths are intrinsic
quantities, corrected for the effective spatial and spectral resolution.
Figure~\ref{fig-maps1} shows the H$\alpha$ linemaps and p-v diagrams along
the kinematic major axis for all of the \bmbx\ galaxies in the sample.
Figure~\ref{fig-maps2} shows the two-dimensional spatial distributions
of the velocity and FWHM velocity width (here uncorrected for the spectral
resolution) of H$\alpha$ for five of the galaxies.
Together with $\rm Q2343-BX610$ (shown in Figure~\ref{fig-bx610}),
these are the ``best cases'' among the sample observed with SINFONI
in terms of the combination of high S/N, large spatial extent and
good spatial resolution, and large velocity gradient.

\subsection{H$\alpha$ Spectra, Sizes, and Morphologies}   \label{Sub-spec}

In most cases, the integrated spectra from the SINFONI data agree
well with the long-slit spectra published by \citet{Erb03}, \citet{Sha04},
and \citet{Ste04}.  These were obtained at lower spectral resolution
($R \sim 1500$) with NIRSPEC at the Keck\,II telescope except for one,
which was obtained at comparable resolution ($R \sim 3500$) with ISAAC at
VLT UT1.  Differences (significant in a few cases only) can be attributed
to difficulties in night sky line subtraction, S/N limitations for extended
sources or broad lines, and slit orientation and positioning that missed
part of the emission for the largest sources.  In addition to H$\alpha$,
we detected the [\ion{N}{2}]\,$\lambda\,6584$ emission line in 12 of the
14 galaxies.  The integrated [\ion{N}{2}]\,$\lambda\,6548$/H$\alpha$ line
flux ratios (hereafter simply [\ion{N}{2}]/H$\alpha$ ratios) are listed in
Table~\ref{tab-res} (see also Figure~\ref{fig-maps1}).

The H$\alpha$ sizes and morphologies among the sample span a fairly
wide range.  The majority of the sources (11/14) are resolved spatially.
The intrinsic $r_{1/2}$ along the major axis range from about 2 to 7~kpc,
with a mean of 4.4~kpc (median of 4.1~kpc).  Three sources may be only
marginally resolved, but the effective spatial PSF for their data is more
uncertain.  In the most extended sources, the H$\alpha$ emission tends to
be irregular and somewhat clumpy.
In the six best cases (Figures~\ref{fig-maps2} and \ref{fig-bx610}),
the H$\alpha$ emission is clearly detected over $\rm \sim 10 - 20~kpc$.
In all of them except $\rm Q1623-BX528$, the brightest H$\alpha$ peak
is located off the geometric center and the H$\alpha$ morphology
exhibits features reminiscent of a tadpole or chain-like shape.

For two of the sources we observed, HST broad-band optical imaging is
available \citep[see Figure 3 of][]{Erb03}.  ACS and earlier WFPC2 data of
$\rm SSA22a-MD41$ reveal a very clumpy rest-frame UV morphology, with bright
knots to the south-west and diffuse faint emission towards the north-east.
The SINFONI linemap shows that the H$\alpha$ emission follows very well the
overall distribution of the rest-frame UV light (accounting for smearing by
the seeing in our ground-based data).  $\rm Q1623-BX376a$, in contrast, is one
of the most compact in H$\alpha$ among our sample.  WFPC2 imaging indicates it
is also compact in the rest-frame UV but it lies at the boundary between two
detectors, making it difficult to assess its detailed spatial distribution.
In these two cases for which we can make the comparison, the spatial
distribution of the H$\alpha$ and rest-frame UV emission is overall
very similar.

The large H$\alpha$ sizes in several of the \bmbx\ sources observed
with SINFONI may not be surprising in view of our selection criteria,
which emphasized (but were not strictly restricted to) bright and/or
extended targets as indicated by previous long-slit spectroscopy
\citep{Erb03, Erb06b, Sha04, Ste04}.  The typical half-light radius
of \bx\ galaxies is $\rm \approx 2~kpc$ \citep[e.g.][]{Sha04, Erb04}.
It is perhaps more interesting that most of the extended sources in our
sample do not appear regular or centrally concentrated in H$\alpha$ emission.
From ACS imaging, \citet{Erb04} also found a prevalence of irregular and
clumpy morphologies peaking off-center in the rest-frame UV for a sample
of 13 \bx\ galaxies, which were selected primarily to be elongated among
their full spectroscopically-confirmed sample in the GOODS-North field.

One may consider these characteristics in light of the results of
\citet*{Elm04a} and \citet*{Elm04b}.  These authors found from ACS
imaging that the majority of faint blue galaxies at moderate to high redshift
($z \sim 0.5 - 2$) appear as ``chains,'' ``tadpoles,'' and ``clump-clusters''
and exhibit a lack of central bulges \citep*[see also, e.g.,][]{Cow95}.
The analogy with the UV-selected \bmbx\ galaxies is instructive, although
one must keep in mind the differences in selection and in redshift range
in making this comparison.  In addition, the rest-frame UV emission may
not trace well the bulk of the stellar population and could miss a bulge
component.  In that respect, the recent study by Toft and coworkers may
shed some new light (S. Toft \etal 2006, in preparation).  These authors
investigated the morphological variations between the rest-frame UV and
optical of $z \geq 2$ galaxies using deep HST ACS and NICMOS imaging at
optical and NIR wavelengths.  They found that the majority of their sources
with blue rest-frame optical colours (as defined by observed NIR $J - K < 2.3$
colours) show little change in morphology, including those that are clumpy
and irregular.  In contrast, their red sources typically become more regular
and often bulge-dominated in the rest-frame optical.
NIR $J$- and $K$-band photometry exists for eight sources
in our sample \citep{Erb06b} and all but two have $J - K < 2.3$.
In view of this, and given that \bmbx\ galaxies have in general $J - K < 2.3$
colours \citep[e.g.][]{Red05}, the clumpy and asymmetric morphologies in the
SINFONI H$\alpha$ linemaps and in HST optical imaging may reflect intrinsic
features of the overall distribution of the stellar population with a lack
of prominent central bulges.
We return to this point in \S~\ref{Sect-case} and \ref{Sect-disc}.

\subsection{H$\alpha$ Kinematics}    \label{Sub-kin}

From the integrated H$\alpha$ line widths
(corrected for instrumental resolution), we derive a mean and rms
velocity dispersion of $\rm 130 \pm 50~km\,s^{-1}$ for our \bmbx\ galaxies,
with a large range from about 70 to $\rm 240~km\,s^{-1}$.  This is in good
agreement with the results of \citet{Erb03, Erb04}, \citet{Sha04}, and
D. Erb \etal (2006, in preparation) from long-slit spectroscopy
of various \bmbx\ samples.  For nine galaxies, we find significant
spatially-resolved velocity gradients (see Figure~\ref{fig-maps1}).
At first glance, this may not appear too surprising.  Indeed, four of the
sources were preferentially chosen as having tilted H$\alpha$ emission in
long-slit spectra.  In those cases, the new SINFONI data add two-dimensional
information and higher spatial resolution.  For the remaining five sources,
however, the existing long-slit data did not show any significant velocity
gradient.

For the six best spatially-resolved (since most extended) galaxies
with two-dimensional kinematic information, the velocity fields appear
smooth and ordered (see Figures~\ref{fig-maps2} and \ref{fig-bx610}).
Similar to local star-forming galaxies \citep[e.g.,][]{Elm98, Gor00},
a smooth kinematic structure
--- despite a clumpy and asymmetric H$\alpha$ light distribution ---
probably is indicative of ordered orbital motions in a disk or a merger.
With the exception of a few objects, our modest linear resolution
($\rm \sim 4~kpc$) compared to the source size does not allow us to uniquely
distinguish between orbital motions of two galaxies in a (widely separated)
merger or rotation in a single disk.  For the purpose of mass estimates using
the virial theorem discussed below, the difference between the merger and
single disk cases, however, is unlikely to be very different.
In analogy with the ultra-luminous infrared galaxies in the local universe
\citep[e.g.,][]{Dow98, Gen01}, the gas in an advanced merger probably lies
mostly in a single plane given that it dissipates efficiently on order of a
dynamical timescale and settles in a rotating disk- or ring-like structure
of moderately large scale at the center of the gravitational potential.
For the same observed parameters and orientation of the orbital plane
relative to the line of sight, a merger model yields a factor $\leq 2$
times larger dynamical mass than a rotating disk model \citep*{Hei85}.

The fact that seven of the nine galaxies with spatially-resolved velocity
structures exhibit the largest velocity gradients approximately along the
morphological major axis supports a rotating disk interpretation.
This contrasts with the findings of \citet{Erb04} based on long-slit
spectroscopy of mostly elongated \bx\ sources in the GOODS-North field.
The reason for this difference is unclear; it could possibly reflect in
part a combination of selection biases for both samples and the comparatively
lower velocity resolution ($\rm \sim 210~km\,s^{-1}$) and angular resolution
(mostly $0\farcs 7 - 0\farcs 9$) for the \citet{Erb04} long-slit data.
Stronger evidence for rotating disks is that the velocity gradients
along the major axis of $\rm SSA22a-MD41$, $\rm Q2343-BX389$, and
$\rm Q2343-BX610$ flatten at the largest radii probed by the SINFONI
H$\alpha$ linemaps, suggestive of the flat part of a disk rotation
curve (see Figure~\ref{fig-velprof}).  These three sources also show a
maximum in FWHM velocity width at their geometric and kinematic center
(and not at the brightest H$\alpha$ emission peak offset from the center;
Figures~\ref{fig-maps2} and \ref{fig-bx610}), which is a key property of
rotating disks.  Finally, the velocity field of $\rm Q2343-BX610$ appears
to exhibit a two-dimensional ``spider diagram'' (Figure~\ref{fig-spider}),
a compelling signature of a classical rotating disk \citep{vdK78}. 
We discuss these three galaxies in more detail in \S~\ref{Sect-case}.

In contrast, the reversal in velocities observed for $\rm Q1623-BX528$
(Figure~\ref{fig-maps2}) could be indicative of a counter-rotating merger
\citep[e.g.,][]{Mih98, Tec00, Gen01}.  The source $\rm Q1623-BX663$ exhibits
a smooth velocity field despite its complex surface brightness distribution,
which could be either due to a relatively face-on disk or a merger.
The largest FWHM velocity width ($\rm \sim 500~km\,s^{-1}$) occurs off-center,
towards the prominent north-eastern H$\alpha$ peak.  The latter characteristic
is rather more consistent with the source being a merger remnant or disturbed
spiral with two long tidal tails, similar to the local analog UGC\,10214
\citep[the ``Tadpole Galaxy,'' e.g.,][]{Tra03}.  The larger velocity width
towards the H$\alpha$ peak, or ``head,'' of $\rm Q1623-BX663$ may then perhaps
be attributed to an AGN located there, consistent with AGN spectral features
present in its rest-frame UV spectrum (A. Shapley, private communication).

Keeping in mind that higher resolution observations will likely reveal
a range of kinematic properties, which are perhaps very complex in any
given source, the evidence for several of our best sources nevertheless 
encourages us to analyze the entire sample in terms of a rotating disk toy
model to derive first order estimates of circular velocities and dynamical
masses.  We considered simple models consisting of azimuthally symmetric
rotating disks, in which the input light distribution is assumed to trace
the distribution of the mass surface density.  Other input parameters include
the total mass, the inclination angle $i$ of the normal to the disk's plane
with respect to the line of sight, and the vertical ($z$-axis) thickness of
the disk from which the $z$ component of the velocity dispersion is computed
under the assumption of hydrostatic equilibrium in the limit of a thin and
very extended disk.  Furthermore, the model allows for an additional
one-dimensional, local isotropic velocity dispersion component $\sigma_{0}$.
The beam smearing due to seeing is taken into account by convolution of
the inclined model with a two-dimensional Gaussian PSF of specified FWHM.

We used these models to estimate the projected circular velocity
at turnover, $v_{\rm c} \sin(i)$, from our measurements of $v_{\rm r}$
(half the maximum observed velocity gradient) and $\Delta v$ (the FWHM
velocity width) as follows.  For rotating disk models that account for beam
smearing and with a range of sizes and of local one-dimensional velocity
dispersions appropriate for the \bmbx\ galaxies in our sample
\footnote{For the purpose of this analysis, we assumed Gaussian light
profiles but similar results are derived with exponential disks or
ring-like distributions.},
the ratio $v_{\rm c}\sin(i) / v_{\rm r}$ is $\sim 1.3$.  The H$\alpha$
emission may preferentially trace the central parts of the galaxy inside the
turnover radius, so that the observed velocity gradient could underestimate the
intrinsic velocity difference at turnover.  This ``velocity gradient'' method
thus yields $v_{\rm c}^{\rm velgrad}\sin(i)$ values that may represent lower
limits to the actual $v_{\rm c}\sin(i)$.  In the same framework and with
similar ranges of parameters, $v_{\rm c}\sin(i) / \Delta v$ is $\sim 0.42$.
If there is any significant line broadening other than due to the projection
of the rotation and random motions related to the disk thickness (e.g., caused
by outflows/winds or streaming motions), this ``line width'' method yields
$v_{\rm c}^{\rm width}\sin(i)$ estimates that may constitute upper limits
to the intrinsic $v_{\rm c}\sin(i)$.
In agreement with these expectations, the velocity gradient method leads
to values on average 0.8 times those obtained with the line width method.
We find the best agreement for the six best resolved galaxies, for which
the mean $v_{\rm c}^{\rm velgrad} / v_{\rm c}^{\rm width}$ is 1.1.  We
adopted the average of the values from the two methods as final estimates.
For the 14 \bmbx\ galaxies studied here, $v_{\rm c}\sin(i)$ ranges from 60
to $\rm 250~km\,s^{-1}$.  Assuming that the sources have random inclinations
and correcting for $1 / \langle \sin(i) \rangle = 1.57$ gives an average
and rms $\langle v_{\rm c} \rangle = 180 \pm 90~{\rm km\,s^{-1}}$.
Table~\ref{tab-res} lists the dynamical masses enclosed
within the H$\alpha$ half-light radius, calculated as
$M_{\rm dyn}\,\sin^{2}(i) = [(v_{\rm c}\,\sin(i))^{2}\,r_{1/2}] / G$,
where $G$ is the gravitational constant.  For the three galaxies for
which more detailed kinematic modeling is possible (\S~\ref{Sect-case}),
the dynamical masses estimated from the simple recipes above are close
to (a factor of 0.82 on average) those derived from the modeling.

\subsection{Additional Considerations}   \label{Sub-cons}

The new SINFONI data represent a significant step forward in our ability
to analyze the kinematics of high redshift galaxies because of the full
two-dimensional spatial mapping at good angular and spectral resolution.
A major advantage of integral field spectroscopy is that no a-priori
assumption about the kinematic major axis needs to be made.
Nevertheless, the interpretation of velocity profiles and maps obviously
requires some caution for observations of distant systems where uncertain
inclination and beam smearing due to seeing can play an important role
\citep[see, e.g., the discussion by][]{Erb04}.

As stated above, we cannot rule out a merger interpretation for our
\bmbx\ sources, except possibly for a few of them, mainly because of spatial
resolution limitations.  The simulations of integral field spectroscopy data
by \citet*{Law06} indicate that while it may still be difficult to distinguish
between rotating disks and two-component mergers for faint and/or compact high
redshift systems, such a distinction is however possible for extended sources
if the data have sufficient angular and spectral resolution, and sufficient
S/N.  This may be the case for the three best cases of our sample (discussed
in \S~\ref{Sect-case}), especially $\rm Q2343-BX610$ for which the high
quality data reveal key kinematic signatures favouring a rotating disk
scenario.

Other factors also need to be considered.  For instance, outflows and winds
may affect the H$\alpha$ kinematics.  In local starburst galaxies with
``superwinds,'' the outflowing gas does not dominate the emission line
kinematics along the major axis \citep{Leh96a}, neither does it dominate the
total high surface brightness H$\alpha$ line emission \citep{Leh95, Leh96b}.
Even in local infrared luminous starburst galaxies with large extinction,
\citet{Arm90} found that the extended emission line gas outside of a few
kiloparsecs from the nucleus accounts for $< 25\%$ of the total H$\alpha$
emission.  For the luminous $z = 2.56$ dust-rich starburst galaxy
$\rm SMM\,J14011+0252$, our SINFONI data \citep[see][]{Tec04} indicate
an outflow component but its contribution to the integrated H$\alpha$
emission is $< 10\%$.  Differential extinction therefore does not seem
in general to play a major role in the relative contribution of outflows
compared to gas photoionized directly by the starburst.
There is also little evidence for a significant amount of shock-heated gas
in \bmbx\ galaxies (such as high [\ion{N}{2}]/H$\alpha$ ratios, as discussed
by \citealt{Leh96a}; see Table~\ref{tab-res} and \citealt{Erb03, Sha04};
D. Erb \etal 2006, in preparation).
It is thus unlikely that mechanically heated gas plays a significant role
in determining the kinematics along the major axis of \bmbx\ objects.

\section{CASE STUDIES: ROTATION CURVES OUT TO $\rm \ga 10~KPC$ AND
                       METALLICITY AND STELLAR POPULATION GRADIENTS}
                       \label{Sect-case}

For $\rm SSA22a-MD41$, $\rm Q2343-BX389$, and $\rm Q2343-BX610$, we carried
out significantly deeper H$\alpha$ observations than for the rest of the
sample (integration times of 8, 5, and 6\,hr, respectively).  This has
enabled us to study their kinematics on radial scales $\rm \ga 10~kpc$
and determine a detailed velocity and mass model out to the asymptotic
part of the rotation curve.  Figure~\ref{fig-velprof} shows the velocity
profile along the kinematic major axis derived from the SINFONI data of
these three galaxies.  For $\rm Q2343-BX610$, the data further allowed us
to investigate spatial variations of the metallicity and stellar population.

We applied the axisymmetric rotating disk models described in
\S~\ref{Sub-kin} to fit the H$\alpha$ kinematics of the galaxies.
From the models, we then determined the following key properties:
the inclination angle $i$, the total ($r > 10~{\rm kpc}$) dynamical
mass $M_{\rm dyn}$ and the dynamical mass within the half-light radius,
the circular velocity at turnover radius $v_{\rm c}$ and at half-light
radius, the vertical scale height of the disk $z_{\rm 1/e}$, and the ratio
$v_{\rm c} / \sigma$ at turnover radius (where $\rm \sigma$ is the velocity
dispersion in the $z$-direction).  We explored a range of model parameters
and fit the data by tuning the main parameters interactively.  Our principal
criterion for the goodness-of-fit was that the models reproduce the observed
radial profile of the velocity along the major axis (Figure~\ref{fig-velprof})
as well as that of the FWHM velocity width.  In addition, the models have
to account for the major axis size (at FWHM or 20 percentile) and the
minor to major axis aspect ratio.   We otherwise gave little weight to the
details of the spatial distribution of the H$\alpha$ emission since it can
be significantly affected, for instance, by extinction and by the actual
distribution of \ion{H}{2} region complexes within the galaxy.  These,
however, do not influence the overall kinematics traced by H$\alpha$.

We explored exponential disk models as well as models where the light/mass
distribution is in form of a ring, as is perhaps suggested by the observed
H$\alpha$ morphologies, which peak off-center.  We found that both types of
models can provide good fits to the kinematic data.  In terms of the surface
brightness distributions, however, the exponential models are generally too
centrally peaked and also have outer wings that are too large.  Ring models
do better in this respect but require, in addition, a more extended component
to reproduce the H$\alpha$ emission at $r > 1^{\prime\prime}$ that is observed
in all three galaxies.  For the exponential disk models, a better fit is
obtained if the disk has a central hole (of radius $0\farcs 1 - 0\farcs 2$),
which is conceptually similar to a ring model.  Exponential disks with scale
lengths of $0\farcs 3 - 0\farcs 5$, or rings with radii and thicknesses
$R \approx \Delta R \approx 0\farcs 3 - 0\farcs 5$ combined with a more
extended component (e.g., a Gaussian of FWHM $\approx 1\farcs 3$), both
match adequately the overall surface brightness distribution of the
three galaxies.

We note that of the 14 \bmbx\ galaxies we observed with SINFONI,
only a very few (the most convincing case is $\rm Q2343-BX389$) exhibit a
classical ``double-horn'' integrated spectral profile, which is characteristic
of motions in a thin azimuthally symmetric plane, as for rotating thin disks
and some mergers.  For such models to match the data satisfactorily, it is
thus required that large random motions throughout the source wash out the
double-horn pattern.  In the case of rotating disks, another contributing
factor may be a slowly rising rotation curve.  The inevitable consequence of
such random motions is that the planar structure must be geometrically thick.

\subsection{$\rm SSA22a-MD41$}    \label{Sub-MD41}

$\rm SSA22a-MD41$ is one of the largest sources among our \bmbx\ sample, with
H$\alpha$ emission detected over a projected diameter of 2\farcs 9 (24~kpc).
As discussed in \S~\ref{Sub-spec}, the H$\alpha$ light distribution follows
well the rest-frame UV morphology and relative intensities seen in deep HST
imaging.  In the context of the \citet{Elm04a, Elm04b} analysis,
$\rm SSA22a-MD41$ resembles a ``tadpole'' system with two bright knots
dominating one side of a clumpy disk or ring embedded in a more diffuse
emission.  There is no evidence for a central bulge.

$\rm SSA22a-MD41$ has the third largest projected velocity gradient
in our sample, $v_{\rm r} \approx 140~{\rm km\,s^{-1}}$, first detected
in long-slit data by \citet{Erb03}.  The observed maximum in velocity
dispersion near the geometric and kinematic center of the source is
$\approx 110~{\rm km\,s^{-1}}$.  The two-dimensional velocity field is in
excellent agreement with expectations for a rotating disk/ring model, with a
smooth velocity gradient that is steepest approximately along the morphological
major axis (Figure~\ref{fig-maps2}).  The most striking kinematic feature is
the flattening of the velocity gradient at the largest radii where we still
detect H$\alpha$ emission, $r \approx 1\farcs 5$ ($\rm \approx 12~kpc$) from
the geometric center (Figure~\ref{fig-velprof}).  This feature is consistent
with the asymptotically flat part of a classical rotation curve.

The H$\alpha$ kinematics and light distribution are very well fit
by the rotating disk models described above.  The total dynamical
mass inferred (within the radius of 12~kpc probed by our data) is 
$M_{\rm dyn} = (6.2 \pm 1.5) \times 10^{10}~{\rm M_{\odot}}$, where the
largest source of uncertainty comes from the choice of light distribution
(exponential disk with central hole or ring) and from the uncertainty in
inclination angle ($i = 71^{\circ} \pm 5^{\circ}$).  As would be expected
from their larger spatial wings, exponential disk models give masses near
the upper end of this range, while ring models give masses near the lower end.
About 60\% of this total mass is enclosed within the H$\alpha$ half-light
radius of $r_{1/2} \approx 6.0~{\rm kpc}$.  The intrinsic (i.e., model)
rotation curve rises rapidly to a maximum of $\rm \approx 165~km\,s^{-1}$,
near $r_{1/2}$, with a slow fall-off at larger radii.  In the framework of
our simple constant (exponential) $z$ scale height plus constant additional
isotropic velocity dispersion models, the best fits require a $z$ scale height
of $z_{\rm 1/e} \approx 0\farcs 17$ and a ratio of peak rotational velocity
to total $z$-velocity dispersion of $v_{\rm c} / \sigma = 2.2 \pm 0.5$.
Clearly the disk is geometrically thick and highly turbulent.

\subsection{$\rm Q2343-BX389$}    \label{Sub-bx389}

$\rm Q2343-BX389$ has the largest projected velocity gradient among
our sample, $v_{\rm r} \approx 235~{\rm km\,s^{-1}}$, and H$\alpha$ is
detected over a projected diameter of 2\farcs 7 (22~kpc).  The observed peak
in velocity dispersion near the geometric and kinematic center is very large,
$\approx 180~{\rm km\,s^{-1}}$.  The H$\alpha$ morphology is reminiscent of
``chain'' galaxies \citep{Elm04a, Elm04b}, with three main knots forming a
linear structure of half-light radius $r_{1/2} \approx 7~{\rm kpc}$.
Both the two-dimensional velocity field and the radial velocity profile are
as expected for a nearly edge-on rotating disk (Figures~\ref{fig-maps2} and
\ref{fig-velprof}).  The velocity gradient is steepest along the morphological
major axis and flattens at the largest radii probed by the data,
$r \approx 1\farcs 2$ ($\rm \approx 10~kpc$).

The kinematic features and the overall light distribution of
H$\alpha$ are well reproduced by disk models with an inclination angle
of $i = 80^{\circ} \pm 5^{\circ}$.  The total dynamical mass derived from
the models is $M_{\rm dyn} = (1.95 \pm 0.5) \times 10^{11}~{\rm M_{\odot}}$,
of which about 70\% is enclosed within $r_{1/2} = 7.0~{\rm kpc}$.  Again,
uncertainties are dominated by those of the light distribution and inclination
but in this case, the difference between exponential disk and ring models is
less than for $\rm SSA22a-MD41$.  $\rm Q2343-BX389$ has the largest derived
dynamical mass of our sample.   The intrinsic rotation curve reaches a
maximum of $\rm \approx 300~km\,s^{-1}$, near the half-light radius, and
decreases slowly at larger radii.  Good fits also require the combination
of a substantial $z$ scale height ($z_{\rm 1/e} \approx 0\farcs 14$) and
corresponding $z$-velocity dispersion, as well as an additional isotropic
velocity dispersion of $\rm 85~km\,s^{-1}$.
This implies $v_{\rm c} / \sigma = 3.3 \pm 0.7$.

\subsection{$\rm Q2343-BX610$}    \label{Sub-bx610}

$\rm Q2343-BX610$ has the second largest projected size and
velocity gradient among our sample, with H$\alpha$ detected over
2\farcs 4 (20~kpc) and $v_{\rm r} \approx 165~{\rm km\,s^{-1}}$. 
The half-light radius is $r_{1/2} \approx 5.4~{\rm kpc}$.
The observed peak in velocity dispersion near the geometric and kinematic
center is $\approx 160~{\rm km\,s^{-1}}$.  $\rm Q2343-BX610$ is remarkably
similar to $\rm SSA22a-MD41$ in its morphological (``tadpole'' shape) and
kinematic features.  Its observed rotation curve also appears to flatten
at radii $\geq 1^{\prime\prime}$ ($\rm \geq 8~kpc$) from the geometric
center (Figure~\ref{fig-velprof}).  This galaxy constitutes our best data
set in terms of S/N and kinematic detail.  This enabled us to map the
two-dimensional distribution of velocity and FWHM velocity width with
good significance over a large fraction of the galaxy, encompassing
about 2~square arcsec (or 16 spatial resolution elements).

The two-dimensional $v$ and $\Delta v$ distributions
are in excellent agreement with expectations for a moderately inclined
rotating disk.  The velocity map shows the telltale ``spider diagram''
pattern (Figure~\ref{fig-spider}) and the velocity dispersion clearly
peaks at the geometric center of the source (Figure~\ref{fig-bx610}), both
basic characteristics of rotating disks \citep{vdK78}.  For this galaxy,
we also detected the continuum in the $K$-band, tracing the rest-frame
optical continuum emission from stars that likely dominate the mass in
the central regions.  The spatial coincidence of the continuum emission
peak and of the kinematic center from the velocity map lends further
support for the rotating disk scenario.

Rotating disk models (either exponential disks with a central hole or rings)
provide excellent fits to the data of $\rm Q2343-BX610$, including in this
case the two-dimensional velocity field of Figure~\ref{fig-spider}.
The axis ratio of the emission indicates an inclination angle of
$i = 63^{\circ} \pm 5^{\circ}$.  The total dynamical mass of this disk is
$M_{\rm dyn} = (9.2 \pm 1.4) \times 10^{10}~{\rm M_{\odot}}$, of which 75\%
is enclosed within the half-light radius.  The difference between masses from
exponential disks and from ring models is small.  The peak rotation velocity
is $v_{\rm c} = 225~{\rm km\,s^{-1}}$, near the half-light radius.
As for the other two sources, our simple models require a substantial $z$
scale height ($z_{\rm 1/e} \approx 0\farcs 16$) and associated $z$ velocity
dispersion, and an additional large isotropic velocity dispersion
($\rm 77~km\,s^{-1}$).
The inferred $v_{\rm c} / \sigma$ ratio is $2.4 \pm 0.6$.

\subsection{The Dynamical Center, $K$-band Surface Brightness,
            and Metallicity Gradient in $\rm Q2343-BX610$}    
            \label{Sub-bx610_2}

$\rm Q2343-BX610$ has the brightest $K$-band magnitude and the highest
[\ion{N}{2}]/H$\alpha$ ratio of the \bmbx\ sample studied here, with
total $K = 19.2~{\rm mag}$ (D. Erb \etal 2006, in preparation)
and [\ion{N}{2}]/H$\alpha = 0.40$ (Table~\ref{tab-res}).  Its brightness
has allowed us to investigate the relative spatial distribution of the line
and continuum emission, and their relation to the emission line kinematics.

Figure~\ref{fig-bx610} shows the detailed properties of $\rm Q2343-BX610$,
including maps of the velocity-integrated H$\alpha$ and [\ion{N}{2}] line flux,
the [\ion{N}{2}]/H$\alpha$ ratio, and the line-free $K$-band continuum tracing
the rest-frame optical continuum.  The [\ion{N}{2}]/H$\alpha$ ratio clearly
varies spatially, reaching a maximum near the geometric center of the
continuum emission.  The peaks of [\ion{N}{2}]/H$\alpha$ ratio and $K$-band
continuum emission are within $\sim 0\farcs 2$ of each other.  While there
is H$\alpha$ emission detected at this position, there clearly is no local
maximum.  The [\ion{N}{2}]/H$\alpha$ ratio measured in a 0\farcs 5--diameter
circular aperture centered where the ratio peaks is $0.55 \pm 0.12$.
Within an annulus of inner diameter 0\farcs 5 and outer diameter 1\farcs 5,
we estimate an average ratio of $0.38 \pm 0.04$ for the rest of the galaxy.
The maximum value of [\ion{N}{2}]/H$\alpha$ is rather high but similar
ratios have been observed in \ion{H}{2} regions of nearby spiral galaxies
\citep[e.g.,][]{Ryd95}.  Furthermore, there is no evidence for the
presence of an AGN in a deep optical spectrum obtained with LRIS at the
Keck~I telescope (D. Erb \etal 2006, in preparation).
In support of this result, our SINFONI $H$-band data give an integrated
ratio of [\ion{O}{3}]\,$\lambda 5007$/H$\beta = 1.5 \pm 0.4$, which, combined
with the integrated [\ion{N}{2}]/H$\alpha$ ratio, places $\rm Q2343-BX610$
in a region still populated by non-AGN galaxies and consistent with normal 
star-forming galaxies in the [\ion{O}{3}]\,$\lambda 5007$/H$\beta$ versus
[\ion{N}{2}]\,$\lambda 6584$/H$\alpha$ diagnostic diagram
\citep[e.g.,][]{Leh94, Kew01, Kau03}.
We conclude that the spatial variations of the [\ion{N}{2}]/H$\alpha$ ratio
and continuum emission are most likely due to gradients in metallicity and
stellar populations.

The relationship between oxygen abundance and [\ion{N}{2}]/H$\alpha$
flux ratio proposed by \citet[][``N2'' index]{Pet04}, 
\begin{equation}
12 + \log({\rm O/H}) = 8.90 + 0.57 \times \log({\rm [NII]}/{\rm H}\alpha),
\label{Eq-OHabund}
\end{equation}
implies $12 + \log({\rm O/H}) = 8.75 \pm 0.04$ at the location of the
[\ion{N}{2}]/H$\alpha$ peak and $8.66\pm 0.03$ for the rest of the galaxy.
For an assumed solar value of $12 + \log({\rm O/H}) = 8.65$
\citep[e.g.][]{Asp04}, we infer an oxygen abundance about 1.25 times
solar towards the center of $\rm Q2343-BX610$ and approximately solar for
the rest of the galaxy.  Since the [\ion{N}{2}] emission, and thus the
[\ion{N}{2}]/H$\alpha$ ratio, tends to saturate at solar metallicity and
above \citep[see][and references therein]{Pet04}, the abundance gradient
in $\rm Q2343-BX610$ could be even higher than implied by the application
of equation~(\ref{Eq-OHabund}).
We also used our additional $J$- and $H$-band data for this source
to estimate the global abundance by applying the $R_{23}$ method
\citep{Pag79} as well as the ``O3N2'' index calibration of \citet{Pet04},
with appropriate corrections for extinction and ionization effects
\citep[as in, e.g.,][]{Tec04}.  Integrated over the source, the $R_{23}$
method implies $12 + \log({\rm O/H}) = 8.54$ and the O3N2 index yields
$12 + \log({\rm O/H}) = 8.55$, both of which are consistent with the
estimate based on the N2 index
\footnote{The S/N ratio of the $J$- and $H$-band data prevents us from
measuring reliably spatial gradients in line ratios and metallicity
indicators.}.
The observed velocity of the higher metallicity region indicates that
it is located near the kinematic center of the galaxy.  In addition,
the relative spatial distributions of the $K$-band and H$\alpha$ emission
imply a significant decrease in the H$\alpha$ equivalent width (EW) towards
the center.  The rest-frame EWs within a 0\farcs 5--diameter aperture
centered on the $K$-band peak and integrated over the galaxy are
$W_{\rm H\alpha}^{\rm rest} = 37 \pm 3$~\AA\ and $53 \pm 2$~\AA,
respectively, suggestive of an increase in the age of the stellar
population towards the $K$-band continuum peak
\citep[see, e.g., the model predictions from STARBURST99;][]{Lei99}.

The SINFONI data of $\rm Q2343-BX610$ thus indicate that the $K$-band
continuum peaks very close to the kinematic center of the galaxy, while the
H$\alpha$ emission, tracing intense star-forming sites, peaks off-center.
The observed gradients in [\ion{N}{2}]/H$\alpha$ ratio and H$\alpha$ EW
suggest significant spatial variations in metallicity and stellar ages,
with higher metallicities and older ages towards the dynamical center.
Overall, $\rm Q2343-BX610$ appears to be one of the most evolved galaxies
among our sample.  In addition to its high [\ion{N}{2}]/H$\alpha$ and
bright $K$-band flux, it has a red $J - K = 2.24$ colour, very close
to the criterion for selecting $z > 2$ ``Distant Red Galaxies''
\citep[DRGs, with $J - K > 2.3$;][]{Fra03, Dok03},
which are found to be generally more evolved than UV-selected galaxies
at similar redshifts \citep{FS04, Lab05}.  The overlap between \bmbx\
galaxies and DRGs is small at $K < 21~{\rm mag}$ \citep{Red05}, but
the metallicity estimates for $\rm Q2343-BX610$ are consistent with
the properties both of DRGs \citep{Dok04} as well as of the brighter
part of the \bmbx\ population \citep[$K \la 20.5$;][]{Sha04}.

\subsection{Mass-Metallicity Relation}   \label{Sub-massmetall}

\citet{Sha04} discussed the first evidence for (super-)solar metallicities
and the possible existence of a luminosity(mass)-metallicity relationship
for \bmbx\ galaxies.  The measurements of the [\ion{N}{2}]/H$\alpha$ ratio
and the dynamical mass estimates inferred in \S~\ref{Sub-kin} for our 14
\bmbx\ galaxies (see Table~\ref{tab-res}) show considerable scatter,
although the two objects with the largest dynamical masses have higher
than average [\ion{N}{2}]/H$\alpha$ ratio.
Converting the [\ion{N}{2}]/H$\alpha$ ratios to O/H abundances with the
\citet{Pet04} relation (see eq.~[\ref{Eq-OHabund}]), we can compare our
${\rm O/H}$ and $M_{\rm dyn}$ estimates with the local relationship between
metallicity and stellar mass derived by \citet{Tre04} using the Sloan Digital
Sky Survey.
Here, it must be borne in mind that a scaling of our dynamical masses within
the half-light radius $r_{1/2}$ to total dynamical masses will lead to an
upward correction, and from dynamical to stellar masses will lead to an 
(unkown) downward correction in the mass coordinate.
Nevertheless, the data suggest that, overall, our \bmbx\ galaxies lie
at lower metallicity (by roughly 0.4~dex) for a given mass with respect
to the local relationship.  This is consistent with the earlier results of
\citet{Sha04} as well as the recent work by \citet{Erb06a}, who discuss in
detail the stellar mass-metallicity relation at $z \sim 2$ based on a large
\bmbx\ sample (including the majority of our SINFONI targets).
The scatter and uncertainties of our [\ion{N}{2}]/H$\alpha$ ratios,
the small mass range covered by our sample, and possible systematic
differences between abundances derived from different methods
\citep[see, e.g.,][and references therein]{Tre04, Sha04, Erb06a}
make it difficult to assess the presence of a mass-metallicity trend
from our data.

\section{DISCUSSION}   \label{Sect-disc}

While the spatial resolution of our ground-based seeing-limited SINFONI
data is still somewhat coarse, they provide us with powerful new tools
to investigate the properties and possible evolution of high-redshift
galaxies based on their kinematics.  In this section, we explore some
implications of our results.  For simplicity, we exclude the $z \approx 1.41$
BM object of our sample and focus on the 13 $z = 2 - 2.5$ \bx\ objects in
the following analysis.

\subsection{Angular Momentum}   \label{Sub-angmom}

Within the context of the $\Lambda$CDM model, the large velocity gradients
within $\rm \sim 10~kpc$ we observe for the \bx\ galaxies of the SINFONI
sample are derived from the tidal torques generated by merging dark matter
halos \citep{Pee69, Whi84}.  If this hypothesis is correct, then these
gradients are directly linked to the spin and specific angular
momentum of the dark matter halo.  Moreover, the amplitude of the gradients
over large radii imply that the galaxies in our sample have significant
amounts of specific angular momentum.  Combining the inferred sizes and
rotation velocities, we can make simple estimates of the specific angular
momenta of the baryons.  An estimate of the angular momentum depends on
the mass-weighted velocity distribution of the galaxies for which we have
only limited constraints.  Assuming a simple model, the angular momentum
of the baryons can be estimated as $j = \beta\,r_{1/2}\,v_{\rm c}$, where
$\beta$ is a dimensionless number that depends on the geometry and spatial
distribution of the mass.  For a disk-like geometry, values of $\beta$ range
between 0.7 for a constant density disk with outer radius $r_{1/2}$ and 2
for an exponential disk of scale length $r_{1/2}$.  Adopting $\beta = 1$,
the average specific angular momentum for the \bx\ sources of
our sample is then $j_{\rm d}^{\rm BX} \sim 900~{\rm km\,s^{-1}\,kpc}$.
The larger rotation velocities and sizes inferred for $\rm SSA22a-MD41$,
$\rm Q2343-BX610$, and $\rm Q2343-BX389$ imply higher specific angular
momenta of $j_{\rm d}^{\rm BX} \sim 1000 - 2000~{\rm km\,s^{-1}\,kpc}$.
Interestingly, these values are comparable to those of local late-type
spirals \citep[e.g.,][and references therein]{Aba03}.

The theoretical explanation for local disk galaxies having significant 
angular momenta compared to that calculated for their dark matter
halos is that there has been little dissipation and net transfer of
angular momentum between the disk's baryons and the dark matter halo
\citep[e.g.,][]{Aba03}.  How does the specific angular momentum that we
estimate for our sample of \bx\ galaxies compare with that expected for
dark matter halos at $z \sim 2$?  In the context of the discussion by
\citet*[][see also \citealt{Whi91}]{Mo98}, the specific angular momentum
of the halos for our BX sample (with $\langle z \rangle = 2.2$) is roughly
given by:
\begin{equation}
j_{\rm h}^{\rm BX} \approx 10^{2.8}\,
   \lambda_{0.05}\,
   \left(\frac{v_{\rm c}}{\rm 180~km\,s^{-1}}\right)^{2}
   \left(\frac{1+z}{3.2}\right)^{-1.5}\,h_{0.7}^{-1}\,{\rm km\,s^{-1}\,kpc},
\label{Eq-angmom}
\end{equation}
where $\lambda_{0.05} \equiv \lambda / 0.05$ is the normalized spin
parameter of the dark matter halo and where we have assumed $\lambda = 0.05$
since this is approximately the most frequent value in the spin parameter
probability distribution \citep[see][]{Mo98}, and $v_{\rm c}$ is the halo
circular velocity at virialization $r_{\rm h,vir}$ in the classical
spherical collapse model \citep{Whi91, Mo02}.
Due to the variations in the spin parameter and to possible differences
in the mass distribution of individual halos, at constant halo circular
velocity the angular momentum of a dark matter halo can vary by a factor
of a few \citep[see][and references therein]{Mo98}.  This large range and
the uncertainty of our estimates of the specific angular momenta means that
we can only make an approximate comparison with theory.  If we hypothesize
that the circular velocities at turnover of our BX galaxies are representative
of the halo virial circular velocities (as we argue in \S~\ref{Sub-DMhalos}),
then these velocities and the range of estimated $j_{\rm d}^{\rm BX}$ values
imply $j_{\rm d}^{\rm BX}/j_{\rm h}^{\rm BX} \sim 1$.
Estimating this ratio to be of order of unity implies that it is indeed
likely that the baryons at small radii did acquire their specific angular
momentum during the collapse of their parent dark matter halos.  Given the
uncertainties, and more speculatively, this would suggest that the baryons
do not transfer their angular momentum to the dark matter upon collapsing
from the virial radius to the disk scale length of a few kiloparsecs.

\subsection{Constraints on the Properties of Dark Matter Halos}
            \label{Sub-DMhalos}

Given that the specific angular momentum of the dark matter halo appears
to be comparable to that of the baryonic disk for the \bx\ galaxies in 
the sample studied here, it is, in principle, possible to constrain the
properties of their dark matter halos using the observed kinematics.

In their analysis of the spatial distribution of UV-selected
galaxies at $1.4 < z < 3.5$ in 21 disjoint fields with total area of
$\rm 0.81~deg^{2}$, \citet{Ade05a} deduced a correlation length for \bx\
objects with $\langle z \rangle = 2.2$ of $4.2 \pm 0.5\,h^{-1}~{\rm Mpc}$.
Through comparison with $\Lambda$CDM numerical simulations of large-scale
structure, \citeauthor{Ade05a} then inferred that the estimated correlation
length corresponds to halo masses of
$M_{\rm h}^{\rm BX} \sim 10^{11.8} - 10^{12.2}~{\rm M_{\odot}}$.
Similarly, \citet{Ade04, Ade05a} found that the observed comoving number
density of \bx\ objects is consistent with the estimated clustering strength,
meaning there are sufficient halos to host galaxies of these halo masses.

If the circular velocity determined on $\rm \la 12~kpc$ scales from our
\bx\ data is a reasonable estimate of the circular velocity of the halo
the galaxy resides in, we can directly estimate the halo mass using the
basic relationship between halo mass $M_{\rm h}$ and circular velocity at
virialization $v_{\rm c}(r_{\rm h,vir})$ in the classical spherical collapse
model \citep{Whi91, Mo02}.
For our BX sample, this yields an average and rms
\begin{equation}
M_{\rm h}^{\rm BX} = 
   10^{11.7 \pm 0.45}\,
   \left(\frac{v_{\rm c}}{\rm 180~km\,s^{-1}}\right)^{3}\,
   h_{0.7}^{-1}\,
   \left( \frac{1+z}{3.2}\right)^{-1.5}~{\rm M_{\odot }},
\label{Eq-Mhalo}
\end{equation}
For $\rm SSA22a-MD41$, $\rm Q2343-BX610$, and $\rm Q2343-BX389$, the average
$M_{\rm h}$ is about a factor of two higher than for the whole sample.  These
$M_{\rm h}$ estimates based on the kinematics are remarkably close to those
obtained by \citet{Ade04, Ade05a} from the clustering properties and comoving
number density of the \bx\ population.  This supports our hypothesis that the
circular velocities we derived for our \bx\ sample may indeed be similar
to those at the halo virial radius.  The analysis of \citet{Whi91} and
\citet{Mo02} also implies a virial radius for the halos of our \bx\
galaxies of
\begin{equation}
r_{\rm h,vir}^{\rm BX} =
   (77 \pm 12)\,
   \left(\frac{v_{\rm c}}{\rm 180~km\,s^{-1}}\right)\,
   h_{0.7}^{-1}\,
   \left(\frac{1 + z}{3.2}\right)^{-1.5}~{\rm kpc}.
\label{Eq-rhalo}
\end{equation}
For $\rm SSA22a-MD41$, $\rm Q2343-BX610$, and $\rm Q2343-BX389$,
the virial radii are about 30\% larger, or $\rm \sim 100~kpc$.
For these three cases, for which our data show a flattening of the
rotation curve at $r_{\rm max} \sim 10~{\rm kpc}$, we thus infer
$r_{\rm max}/r_{\rm h,vir} \sim 0.1$.
At a radius of $0.1\,r_{\rm h,vir}$, simulations \citep[e.g.][]{Hay04}
show that $v_{\rm c}(r) / v_{\rm c}(r_{\rm h,vir}) \sim 0.9$,
supporting our contention that the circular velocities are a reasonable
representation of the halo virial velocity for our \bx\ objects.
Since these simulations do not include the condensation of baryons at
the center of these halos --- which obviously happened for the galaxies
studied here --- the values of $v_{\rm c}(r)$ inferred from these models
constitute lower limits.  Overall, the simple analysis above implies that
the circular velocities we derived for our \bx\ sample are very likely
good estimates of the true circular velocity of their halos.

We note that in a second paper on the spatial clustering focussing on
$z \sim 2$ \bx\ galaxies in four fields with NIR imaging, \citet{Ade05b}
found a substantially higher correlation length for those brighter than
$K = 20.5~{\rm mag}$ compared to the overall population (by a factor of
about 2.4).  However, as the authors point out, much of the signal leading
to this high correlation length comes from a single field, which may bias
the average result (see also the discussion by \citealt{Ste05} who found a
significant redshift overdensity for \bx\ sources in another of the four
fields used by \citealt{Ade05b}).

The estimates of the halo masses and galaxy dynamical masses of our \bx\
sample imply that only a fraction of the baryons are associated with the
galaxies (see \citealt{Ade05a} for a related discussion).  By combining our
derived dark matter halo masses with the dynamical masses, we can estimate
the amount of the possible total baryonic matter in place in our \bx\ galaxies.
For $\rm SSA22a-MD41$, $\rm Q2343-BX610$, and $\rm Q2343-BX389$, our dynamical
modeling (\S~\ref{Sect-case}) leads to
$M_{\rm dyn}(r_{\rm max}) \approx (0.6 - 2) \times 10^{11}~{\rm M_{\odot}}$.
The halo masses derived above assuming virial equilibrium are about a factor
of 10 higher, i.e., the ratio $M_{\rm dyn}/M_{\rm h} \approx 0.1$.  For
comparison, the currently best value for the fraction of baryonic to total
mass is $\Omega_{\rm b} / \Omega_{\rm m} \sim 0.17$ \citep[e.g.,][]{Spe03}.
Taken literally, our results would suggest that even if the dark matter
content is negligible inside a radius of $r_{\rm max}$, only about half or
less of the baryons associated with the dark matter halo are actually within
$r_{\rm max}$.  For the Milky Way, the fraction of total baryonic to dark
matter mass is $0.08 \pm 0.01$ \citep[e.g.,][]{Car05}.  The similarity between
the Milky Way and the three \bx\ galaxies considered here suggest that at
epochs as early as $z \sim 2$, at least some of the \bx\ galaxies might
already have an ``incorporated mass fraction'' comparable to that of
local spiral galaxies
(with this fraction likely to grow further during subsequent evolution
if \bx\ galaxies evolve into present-day elliptical galaxies as suggested
by their clustering; see \citealt{Ade05a}).

\subsection{Dynamical Evolution}   \label{Sub-dyn_state}

The determination of the ratio of circular velocity $v_{\rm c}$ to 
$z$-velocity dispersion $\sigma$ may provide an interesting new probe of the
dynamical evolution of actively star-forming galaxies at high redshift.
We consider again the three best cases, $\rm SSA22a-MD41$, $\rm Q2343-BX610$,
and $\rm Q2343-BX389$, for which the modeling of \S~\ref{Sect-case} implies
$v_{\rm c}/\sigma $ and $\rm r_{1/2}/z_{1/e}$ of $2 - 4$.  These ratios are
much lower than the value of $\ga 10$ for the Milky Way disk.  We here briefly
explore two possible mechanisms through which the local velocity dispersion
of the gas can be increased in high redshift star-forming galaxies.

In one scenario, the energy from the gravitational potential is converted
into random motions during gas infall onto a disk.  If the conversion is
efficient, $v_{\rm c} / \sigma$ gives a measure of the rate of gas infall
(N. Murray, private communication).  Assuming that random motions are
generated by viscous processes in a rotating disk (as considered, e.g.,
by \citealt{Jog88} and \citealt*{Gam91} for lower velocity dispersions
in the Galactic disk; see also \citealt{Gol78, Fuk84}) and that the
rotational and random motion energies grow at approximately equal rates,
then
\begin{equation}
\sigma / v_{\rm c} \approx 
   \eta\,
   \sqrt{\frac{t_{\rm dyn}}{t_{\rm acc}}},
\label{Eq-tacc}
\end{equation}
since dispersion is generated on a dynamical timescale and rotational
energy increases at the mass accretion rate.  In equation~(\ref{Eq-tacc}),
$t_{\rm dyn} = r / v_{\rm c}$ is the dynamical timescale,
$t_{\rm acc} = M_{\rm gas}/(dM_{\rm gas}/dt)$ is the mass accretion timescale
onto the disk, and $\eta$ is a dimensionless parameter with value near unity.
In this scenario, the values of $v_{\rm c}/\sigma \sim 2 - 4$
for $\rm SSA22a-MD41$, $\rm Q2343-BX610$, and $\rm Q2343-BX389$
translate into $t_{\rm acc}/t_{\rm dyn} \sim 4 - 16$.
Given their $t_{\rm dyn} \sim 50~{\rm Myr}$
at $r_{\rm max} \sim 10~{\rm kpc}$, this implies 
$t_{\rm acc} \sim 200 - 800~{\rm Myr}$
(consistent with scenarios of accretion through cold flows
of baryons from the halo; e.g., \citealt{Dek06}).
The rotating disk picture assumed here as well as for the modeling of
\S~\ref{Sect-case} neglects possible contributions to the observed velocity
dispersion due to non-circular motions, merger-induced flows, or other
perturbations.  Therefore, the $\sigma / v_{\rm c}$ values are very
likely upper limits in the context of equation~(\ref{Eq-tacc}) so that the
accretion timescale estimates represent almost certainly lower limits.
\citet{Erb06b} applied population synthesis models to fit the optical to
NIR (and Spitzer/IRAC mid-IR for a subset) spectral energy distribution (SED)
of a large \bx\ sample including $\rm Q2343-BX610$ and $\rm Q2343-BX389$, for
which they derived stellar ages $\rm \ga 1~Gyr$ (at 95\% confidence), at the
high end of the range of $\rm \sim 100~Myr - 1~Gyr$ typical for \bx\ galaxies
\citep[see also][]{Sha05}.  Interestingly enough, the lower limit for
$t_{\rm acc}$ and the stellar ages are consistent with each other.

An alternative scenario is that of an actively star-forming disk in which
stellar winds, supernova explosions, and radiation pressure are responsible
for setting the velocity dispersion and providing vertical support against
gravity \citep{Tho05}.  If the disk reaches marginal stability in terms
of Toomre's Q-parameter \citep{Too64}, the relationship between
$\sigma / v_{\rm c}$ and the gas fraction $f_{\rm gas}$ can be written as
\begin{equation}
\sigma / v_{\rm c} \approx \left(f_{\rm gas}\,Q\right)/a,
 \label{Eq-Qparam}
\end{equation}
where $a = 2 \pm 1$ is a dimensionless parameter depending on the assumed
distribution of gas and gravitational potential.  In this case, the total
star formation rate $R_{\star}$ may be described by a Schmidt law,
\begin{equation}
R_{\star} = dM_{\rm gas}/dt = 
    \varepsilon\,f_{\rm gas}\,M_{\rm dyn}\left(t_{\rm dyn}\right)^{-1},
\label{Eq-schmidt}
\end{equation}
and the gas exhaustion timescale $t_{\star}$ by star formation is
\begin{equation}
t_{\star} = M_{\rm gas}\,/\left(dM_{\rm gas}/dt\right) =
            t_{\rm dyn}\,/\,\varepsilon.
\label{Eq-tstar}
\end{equation}
Here $\varepsilon$ is the star formation efficiency for converting gas
into stars, which is estimated to be $\sim 0.1$ in the Galactic disk
\citep[e.g.,][]{Elm99}.
For the inferred range of $v_{\rm c} / \sigma \sim 2 - 4$,
equation~(\ref{Eq-Qparam}) and $\rm Q \approx 1$ imply
$f_{\rm gas} \approx 0.5 - 1$.  To be as thick as inferred from our data
the gas disks must be very gas-rich.  For $\varepsilon \approx 0.1$, the gas
exhaustion timescale is about 500~Myr.  This picture, in which the velocity
dispersion in a gas-rich star-forming disk is related to feedback processes
from star formation, thus leads to estimates of gas exhaustion timescales
that are plausibly consistent with the stellar ages as well.

\subsection{Disk Stability and Subsequent Evolution}   \label{Sub-stab}

In view of the above, we now consider the question of the lifetime
of the present dynamical state of the \bx\ galaxies we have observed.
In a merger interpretation with two interacting galaxies within
$\rm \approx 10~kpc$, the system would form a compact (i.e., bulge) merger
remnant within $\approx 10$ dynamical timescales, or $\rm \approx 500~Myr$.
One might suppose that a disk configuration would be more stable.  However,
the disks we appear to be observing in some of the \bx\ galaxies differ
from local Sa/Sb disk galaxies in that they seem very gas-rich and possibly
lack a central bulge.  Such systems are probably dynamically unstable.

\citet{Imm04b} present the results of a chemo-dynamical simulation of the
collapse and fragmentation of a gas-rich galactic disk at high redshift
(one of the possible scenarios for disk evolution and bulge formation
explored by \citealt{Imm04a}).  In this simulation, \citet{Imm04b} start from
a gaseous disk with a mass inflow rate of $\rm 120~{\rm M_{\odot}\,yr^{-1}}$,
growing to a total mass of $\rm 1.2 \times 10^{11}~{\rm M_{\odot}}$ in 1~Gyr.
This disk fragments into massive star-forming clumps, with masses between
0.6 and $\rm 7 \times 10^{9}~{\rm M_{\odot}}$ and
$\sigma_{\rm gas} \sim 30 - 50~{\rm km\,s^{-1}} \sim 0.25\,v_{\rm c}$.
The total mass, the star formation rate (reaching a maximum of
$\rm \sim 220~{\rm M_{\odot}\,yr^{-1}}$ after $\rm \sim 1~Gyr$),
the assembly time, and the knotty appearance of this disk are very
similar to the properties of the clump-cluster galaxies studied by
\citet[][see also \citealt{Elm04a, Elm04b}]{Elm05} and some of the \bx\
galaxies in our sample.  The clumpy fragmenting disk is unstable.  The
star-forming clumps sink to the gravitational center by dynamical friction
against the diffuse underlying disk of gas and form a central bulge after
$\rm \sim 1~Gyr$.  It is interesting to note that this final configuration is
similar to those of high redshift galaxies in the hydrodynamical simulations
by \citet{Aba03}, and that it also resembles that of a spheroid rather than
of a disk galaxy.

In the context of these simulations, it is tempting to speculate that some
of the \bx\ galaxies in our sample are in the mid-stages of the evolution
of such unstable gas-rich disks later evolving into galaxy bulges.
$\rm Q2343-BX610$ may be in a more evolved stage in this sequence, in which
an older and more metal-rich central bulge of substantial mass has already
formed.

\subsection{Comparison Between Stellar and Dynamical Masses}
            \label{Sub-Mdyn_Mstar}

Having estimated dynamical masses for our \bx\ sample, an obvious
test is to compare them with photometric stellar mass estimates.
\citet{Sha05} and \citet{Erb06b} have carried out population synthesis
modeling of the SEDs of large samples of spectroscopically-confirmed \bx\
galaxies.  They found that the average stellar mass in a typical BX galaxy
is $M_{\star} \approx 10^{10.3 \pm 0.5}~{\rm M_{\odot}}$.
\citet{Sha04} further compared the stellar masses and a virial estimate of
the dynamical masses of nine $K \leq 20$ \bx\ objects, which are at the
massive tail ($M_{\star} \sim 10^{11}~{\rm M_{\odot}}$) of the population.
They concluded that the stellar masses exceed the dynamical masses by a
factor of about three for a Salpeter IMF ($dN/dm \propto m^{-2.35}$) 
between $\rm 0.1$ and $100~M_{\odot}$, which is commonly adopted in
studies of high redshift galaxies.

From studies in the Milky Way, it is now clear \citep[e.g.,][]{Sca86}
that the IMF flattens below a few solar masses even if a Salpeter slope
may be applicable at the high mass end. 
With such a flattening at the low mass end, modern versions of appropriate
local IMFs \citep[e.g.][]{Kro01, Cha03} lead to total masses $\sim 1.6$
times lower than the unrealistic $\rm 0.1 - 100~M_{\odot}$ Salpeter IMF.
For a more extreme (but probably also unrealistic) $\rm 1 - 100~M_{\odot}$
Salpeter IMF, total masses are 2.5 times lower.  \citet{Sha04} found that
a \citeauthor{Kro01} or \citeauthor{Cha03} IMF is consistent with their
virial mass estimates derived from the velocity widths for seven \bx\
galaxies.

The 11 \bmbx\ SINFONI targets with available NIR photometry are included
in the sample for which \citet{Erb06b} carried out SED modeling.  For a
\citeauthor{Cha03} IMF, their modeling implies and average and rms stellar
mass of $\langle M_{\star} \rangle \approx 10^{10.8 \pm 0.1}~{\rm M_{\odot}}$.
The dynamical masses we inferred from the velocity gradient and line width
methods in \S~\ref{Sub-kin} for these objects agree well, with a mean and rms
of $\langle M_{\rm dyn} \rangle \approx 10^{10.7 \pm 0.2}~{\rm M_{\odot}}$
within the H$\alpha$ half-light radius.  Taking the results of the more
detailed kinematic modeling of $\rm SSA22a-MD41$, $\rm Q2343-BX610$, and
$\rm Q2343-BX389$ as a guide (\S~\ref{Sect-case}), the ratio of total
mass and mass within $r_{1/2}$ is about 1.5.  With this correction,
$\langle M_{\rm dyn} \rangle / \langle M_{\star} \rangle$ is about
$10^{0.1 \pm 0.3}$.  For $\rm Q2343-BX610$ and $\rm Q2343-BX389$,
the kinematic modeling gives total $M_{\rm dyn}$ of $9.2 \times 10^{10}$
and $\rm 2.0 \times 10^{11}~{\rm M_{\odot}}$, which are factors of 0.4 and
3 times the $M_{\star}$ estimates of \citet{Erb06b}.
These considerations illustrate the uncertainties involved in deriving
photometric stellar masses and dynamical masses.  Nonetheless, the dynamical
masses tend to be overall about equal to the photometric stellar masses with
a \citeauthor{Kro01} or \citeauthor{Cha03} IMF.  Leaving some room for
contributions from gas and dark matter might imply a slight shortfall of
the dynamical masses and would then require a still smaller fraction of
low mass stars or, alternatively, a more top-heavy IMF
\citep[e.g., as proposed by][]{Bau05}.
Our dynamical mass measurements may thus strengthen the arguments
that a $\rm 0.1 - 100~M_{\odot}$ Salpeter IMF is inappropriate for
high redshift galaxies.

\section{Summary}   \label{Sect-conclu}

We have carried out NIR integral field spectroscopy with SINFONI of
a sample of 14 UV-selected \bmbx\ galaxies at $z \sim 2$ to investigate
the morphology and kinematics as traced by their H$\alpha$ line emission.
From our H$\alpha$ linemaps, as well as higher resolution rest-frame UV
imaging from HST for other \bx\ galaxies \citep{Erb04}, it appears that,
at least in part, the \bmbx\ galaxies we have observed are not classical
exponential disks with a central bulge but present rather asymmetric and
clumpy morphologies.  Deep high resolution NIR imaging is obviously highly
desirable to determine the morphologies in rest-frame optical continuum
emission.

We detected spatially-resolved velocity gradients in almost all galaxies
whose integrated size is sufficiently large to allow us to make such a
measurement.  For the six cases for which the data provide high quality
two-dimensional kinematic information, the velocity fields and gradients are
smooth and, in all but one or two cases, match the expectations for a simple
rotating disk.  The steepest velocity gradients for the galaxies in our
sample are generally approximately along the morphological major axis.
For the three best cases among our SINFONI sample, we have obtained
the first convincing asymptotic rotation curves at high redshift,
on radial scales of $\rm \ga 10~kpc$ from the dynamical center.
The dynamics of the \bmbx\ sample studied here thus appear to
be dominated by gravitationally-driven motions.

From these measurements and from kinematic modeling, we inferred dynamical
masses ranging from $\sim 0.5$ to $\rm 25 \times 10^{10}~M_{\odot}$ within
the H$\alpha$ half-light radius.  Hypothesizing that the average $v_{\rm c}$
at a radius of $10~{\rm kpc}$ is an appropriate proxy for the circular
velocity of the dark matter halo implies a virial mass for the typical halo
of a \bmbx\ galaxy in our sample of $10^{11.7 \pm 0.5}~{\rm M_{\odot}}$.
This is in good agreement with estimates based on the clustering and co-moving
space density of the \bmbx\ population \citep{Ade05a}. The kinematic technique
may provide a new, independent method for estimating dark matter halo masses.
From the more detailed kinematic modeling of the three best cases, we inferred
a ratio of peak circular velocity $v_{\rm c}$ to $z$-velocity dispersion
$\sigma$ of $\sim 2 - 4$.  This suggests a mass accretion timescale onto the
disk or, alternatively, a gas exhaustion timescale of $\rm \sim 200 - 800~Myr$,
broadly consistent with the stellar ages.  We found that the specific angular
momenta of the \bmbx\ objects in our sample are similar to those of local
late-type galaxies, and that the specific angular momenta of their baryonic
disks appear to be roughly comparable to those of the dark matter halos
from which they formed.

We presented arguments, based on the properties of the \bmbx\ galaxies
analyzed in this paper, which may suggest an interesting scenario for their
evolution.  Merger-induced or (quasi-)adiabatic gas collapse, whereby the
specific angular momentum of the gas relative to the dark matter halo is
preserved, has its free energy converted into turbulence in the disk, which
is one way to explain the low $v_{\rm c} / \sigma$ we observed.  Gas-rich
disks growing through infall are likely unstable against fragmentation.
The self-gravitating fragments result in clumpy and irregular morphologies
such as seen in the H$\alpha$ linemaps of several of our \bmbx\ sources.
These fragments may transfer angular momentum to infalling gas as they
sink towards the center of the gravitational potential through dynamical
friction, leading to the formation of a central bulge.
Angular momentum transfer in this way may allow the residual gas to
keep its high specific angular momentum, such that it remains comparable
to that of the dark matter halo.  The collapse of the gas would be
relatively inefficient, preserving the relatively low fraction of
baryonic mass that we inferred for our sample.
Although some of these arguments are speculative, they may suggest a
plausible and consistent picture that may explain a number of features
observed for some of the \bmbx\ galaxies studied here.

The results presented in this work apply to the \bmbx\ sample that we have
observed with SINFONI.  Admittedly, the sample may not be representative of
the \bmbx\ population as a whole because the targets were preferentially
selected to be large, or to exhibit significant velocity gradients or large
line widths based on existing long-slit spectroscopy.  Nonetheless, the
SINFONI data we have obtained show the improvement in our ability to
investigate the morphologies and kinematics of high redshift galaxies
afforded by NIR integral field spectroscopy.  It will undoubtedly be
very interesting to extend such studies to other high redshift populations
and further investigate the issues addressed in this work in the future.

\acknowledgments

We wish to thank the ESO staff for helpful and enthusiastic support during
the observations.  We thank the referee for useful comments that helped
improve the paper.  We are also grateful to Norman Murray, Eliot Quataert,
Todd Thompson, Ortwin Gerhard, and Sune Toft for interesting discussions
and insightful comments on various aspects of this work.


\clearpage


\begin{deluxetable}{lcclrl}
\tabletypesize{\small}
\tablecolumns{6}
\tablewidth{360pt}
\tablecaption{Summary of the SINFONI observations of the BM/BX sample
              \label{tab-obs}}
\tablehead{
   \colhead{Source} & 
   \colhead{$z_{\rm H\alpha}$\,\tablenotemark{a}} & 
   \colhead{Band} & 
   \colhead{Scale} & 
   \colhead{$t_{\rm int}$\,\tablenotemark{b}} & 
   \colhead{Run\,\tablenotemark{c}} \\
   \colhead{} & 
   \colhead{} & 
   \colhead{} & 
   \colhead{} & 
   \colhead{(s)} & 
   \colhead{}
}
\startdata
$\rm Q1307-BM1163$ & 1.4101 & $H$ & 0\farcs 125 & 14400 & Mar05 \\
$\rm Q1623-BX376a$ & 2.4087 & $K$ & 0\farcs 125 & 15600 & Mar05, Apr05 \\
$\rm Q1623-BX455$  & 2.4071 & $K$ & 0\farcs 125 & 12000 & Mar05 \\
$\rm Q1623-BX502$  & 2.1555 & $K$ & 0\farcs 05  &  9600 & Apr05 \\
$\rm Q1623-BX528$  & 2.2684 & $K$ & 0\farcs 125 & 13500 & Jul04 \\
$\rm Q1623-BX599$  & 2.3313 & $K$ & 0\farcs 125 &  5400 & Jul04 \\
$\rm Q1623-BX663$  & 2.4335 & $K$ & 0\farcs 125 &  9000 & Jul04 \\
$\rm SSA22a-MD41$  & 2.1710 & $K$ & 0\farcs 125 & 28800 & Nov04, Jun05 \\
$\rm Q2343-BX389$  & 2.1728 & $K$ & 0\farcs 125 & 18000 & Oct05 \\
$\rm Q2343-BX610$  & 2.2102 & $K$ & 0\farcs 125 & 10800 & Jun05 \\
                   &        & $H$ & 0\farcs 125 & 18000 & Aug05, Oct05 \\
                   &        & $J$ & 0\farcs 125 & 14400 & Oct05 \\
$\rm Q2346-BX404$  & 2.0298 & $K$ & 0\farcs 125 &  8100 & Jul04 \\
$\rm Q2346-BX405$  & 2.0308 & $K$ & 0\farcs 125 &  8100 & Jul04 \\
$\rm Q2346-BX416$  & 2.2406 & $K$ & 0\farcs 125 &  7200 & Dec04 \\
$\rm Q2346-BX482$  & 2.2569 & $K$ & 0\farcs 125 &  7200 & Nov04 \\
\enddata
\tablenotetext{a}
{
Systemic vacuum redshift derived from the integrated H$\alpha$ line
emission in the SINFONI data (as described in \S~\ref{Sect-res}).
$\rm Q2348-BX404$ and BX405 form a pair, with angular separation of
3\farcs 6 (or 30~kpc at their redshift).
}
\tablenotetext{b}
{
Total on-source integration time.
}
\tablenotetext{c}
{
SINFONI observing runs where the data were taken:
Jul04: commissioning run on 2004 July 9 to 18;
Nov04: GTO run on 2004 November 29 and 30;
Dec04: GTO run on 2004 December 20 to 22;
Mar05: GTO run on 2005 March 13 to 22;
Apr05: GTO run on 2005 April 4 to 7.
Jun05: GTO run on 2005 June 15 to 17.
Aug05: GTO run on 2005 August 27 to September 4.
Oct05 : GTO run on 2005 October 2 to 12.
}
\end{deluxetable}


\begin{deluxetable}{llccccc}
\tabletypesize{\small}
\tablecolumns{7}
\tablewidth{360pt}
\tablecaption{Properties derived from the SINFONI
              observations of the BM/BX sample.
              \label{tab-res}}
\tablehead{
   \colhead{Source} & 
   \colhead{$z_{\rm H\alpha}$\,\tablenotemark{a}} & 
   \colhead{\protect{[\ion{N}{2}]/H$\alpha$}\,\tablenotemark{b}} &
   \colhead{$r_{1/2}$\,\tablenotemark{c}} &
   \colhead{$\Delta v $\,\tablenotemark{d}} &
   \colhead{$v_{\rm r}$\,\tablenotemark{d}} &
   \colhead{${\sin}^{2}(i) \times M_{\rm dyn}$\,\tablenotemark{e}} \\
   \colhead{} & 
   \colhead{} &
   \colhead{} &
   \colhead{(kpc)} & 
   \colhead{($\rm km\,s^{-1}$)} & 
   \colhead{($\rm km\,s^{-1}$)} &
   \colhead{($10^{10}~M_{\odot}$)} 
}
\startdata
$\rm Q1307-BM1163$   & 1.4101 & $0.22 \pm 0.03$ & 3.1 & 310 &  60 & 0.84 \\
$\rm Q1623-BX376a$   & 2.4087 & $0.15 \pm 0.1 $ & 3.1 & 210 &  30 & 0.33 \\
$\rm Q1623-BX455$    & 2.4071 & $0.14 \pm 0.07$ & 3.5 & 270 &  55 & 0.73 \\
$\rm Q1623-BX502$    & 2.1555 & $< 0.05$        & 2.0 & 170 &  50 & 0.21 \\
$\rm Q1623-BX528$    & 2.2684 & $0.32 \pm 0.05$ & 6.6 & 280 &  65 & 1.7  \\
$\rm Q1623-BX599$    & 2.3313 & $0.24 \pm 0.08$ & 3.4 & 375 &  50 & 1.1  \\
$\rm Q1623-BX663$    & 2.4335 & $0.24 \pm 0.08$ & 5.3 & 360 &  95 & 2.4  \\
$\rm SSA22a-MD41$    & 2.1710 & $0.15 \pm 0.05$ & 6.0 & 300 & 140 & 3.2  \\
$\rm Q2343-BX389$    & 2.1728 & $0.23 \pm 0.07$ & 7.0 & 570 & 235 & 11   \\
$\rm Q2343-BX610$    & 2.2102 & $0.41 \pm 0.02$ & 5.4 & 385 & 165 & 4.6  \\
$\rm Q2346-BX404$    & 2.0298 & $0.20 \pm 0.06$ & 2.8 & 210 &  20 & 0.27 \\
$\rm Q2346-BX405$    & 2.0308 & $< 0.05$        & 4.6 & 170 &  30 & 0.37 \\
$\rm Q2346-BX416$    & 2.2406 & $0.17 \pm 0.05$ & 2.8 & 315 &  70 & 0.84 \\
$\rm Q2346-BX482$    & 2.2569 & $0.19 \pm 0.05$ & 5.8 & 300 & 100 & 2.3  \\
\enddata
\tablenotetext{a}
{
Systemic vacuum redshift derived from the integrated H$\alpha$ line emission.
}
\tablenotetext{b}
{
Ratio of the integrated [\ion{N}{2}]\,$\lambda\,6584$ and H$\alpha$
line fluxes. For $\rm Q1307-BM1163$, $\rm Q1623-BX599$,
and $\rm Q1623-BX663$, there is a broad component to H$\alpha$,
making estimates of the [\ion{N}{2}]/H$\alpha$ ratio more uncertain.
}
\tablenotetext{c}
{
Rest-frame intrinsic half-light radius $r_{1/2}$ along the major axis
of the detected H$\alpha$ light distribution (see \S~\ref{Sect-res}).
The observed HWHM along the morphological major axis of each galaxy was
corrected for a typical NIR seeing with $\rm FWHM = 0\farcs 5$ except
for $\rm Q1623-BX502$, for which AO-assisted observations were taken,
resulting in an effective resolution with $\rm FWHM = 0\farcs 3$.
}
\tablenotetext{d}
{
Intrinsic FWHM velocity width $\Delta v$ and half-total velocity spread
$v_{\rm r} \equiv (v_{\rm max} - v_{\rm min}) / 2$ over the H$\alpha$
emitting regions (see \S~\ref{Sect-res}).  The FWHM line widths for each
galaxy were corrected for the instrumental resolution of
$\approx 80~{\rm km\,s^{-1}}$.
}
\tablenotetext{e}
{
Inferred dynamical mass within $r_{1/2}$ (uncorrected for inclination),
averaged from the ``line width'' and ``velocity gradient'' methods,
as described in \S~\ref{Sect-res}.
}

\end{deluxetable}

\clearpage


\setcounter{figure}{0}

\begin{figure}[p]
\figurenum{1}
\epsscale{1.00}
\plotone{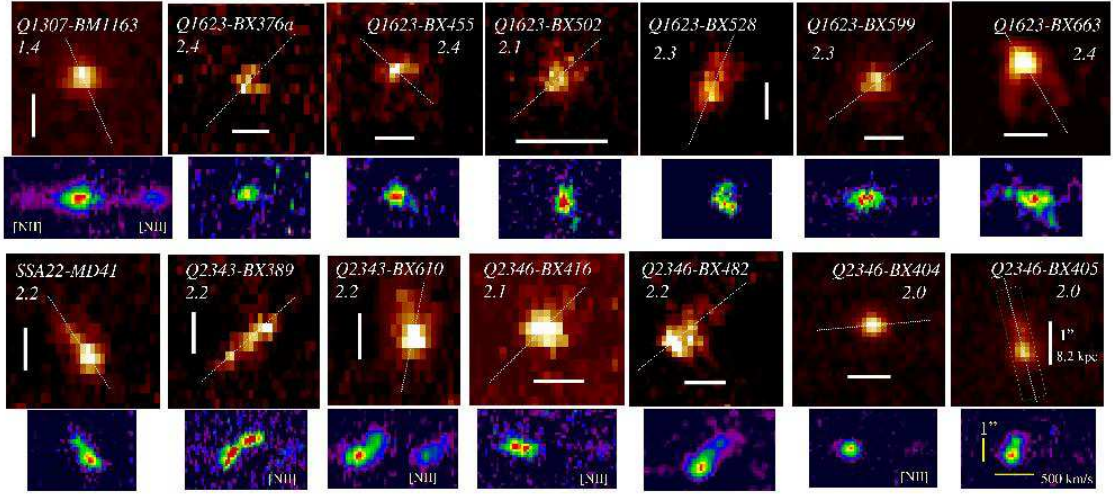}
\vspace{0.5cm}
\caption{
\small
H$\alpha$ morphology and kinematics of the sample
of 14 BM/BX objects observed with SINFONI.
For each galaxy, we show the velocity-integrated H$\alpha$ linemap
({\em top insets\/}) and a position-velocity (p-v) diagram extracted in
a $\approx 0\farcs 5$--wide synthetic slit ({\em bottom insets\/}).
The linemaps are displayed with a linear intensity scale increasing from
black to white, and the p-v diagrams are colour-coded with a linear intensity
scale increasing from magenta to red.  Each linemap is labeled with the
source name and redshift, and the position of the synthetic slit is shown
as well ({\em thin dotted line\/}).  In the inset pair at the bottom right,
the angular and velocity scales for the linemaps and position-velocity
diagrams are marked, and the full synthetic slit width used for all
objects is shown.  The angular size of $1^{\prime\prime}$ is also
indicated in all linemaps ({\em thick solid line\/}), corresponding to
$\rm \approx 8.3~kpc$ at the redshifts of the objects.  For all sources,
the data were obtained with the $\rm 0\farcs 125~pixel^{-1}$ scale,
with the exception of $\rm Q1623-BX502$, which was observed with the
$\rm 0\farcs 05~pixel^{-1}$ scale with AO correction on a nearby star.
In the p-v diagrams, one pixel corresponds to $\rm 0.000195~\mu m$ along the
horizontal (wavelength) axis for the $H$-band data of $\rm Q1307-BM1163$ and
to $\rm 0.000245~\mu m$ for all other sources, observed in the $K$-band.
In all linemaps, north is up and east is to the left.
The objects show a variety of H$\alpha$ morphologies and velocity gradients.
The \protect{[\ion{N}{2}]\,$\lambda\,6584$} emission line is visible and
labeled in the p-v diagrams of $\rm Q1307-BM1163$, $\rm Q2343-BX389$,
$\rm Q2343-BX610$, $\rm Q2346-BX416$, and $\rm Q2346-BX404$.
\label{fig-maps1}
}
\end{figure}

\clearpage

\begin{figure}[p]
\figurenum{2}
\epsscale{0.62}
\plotone{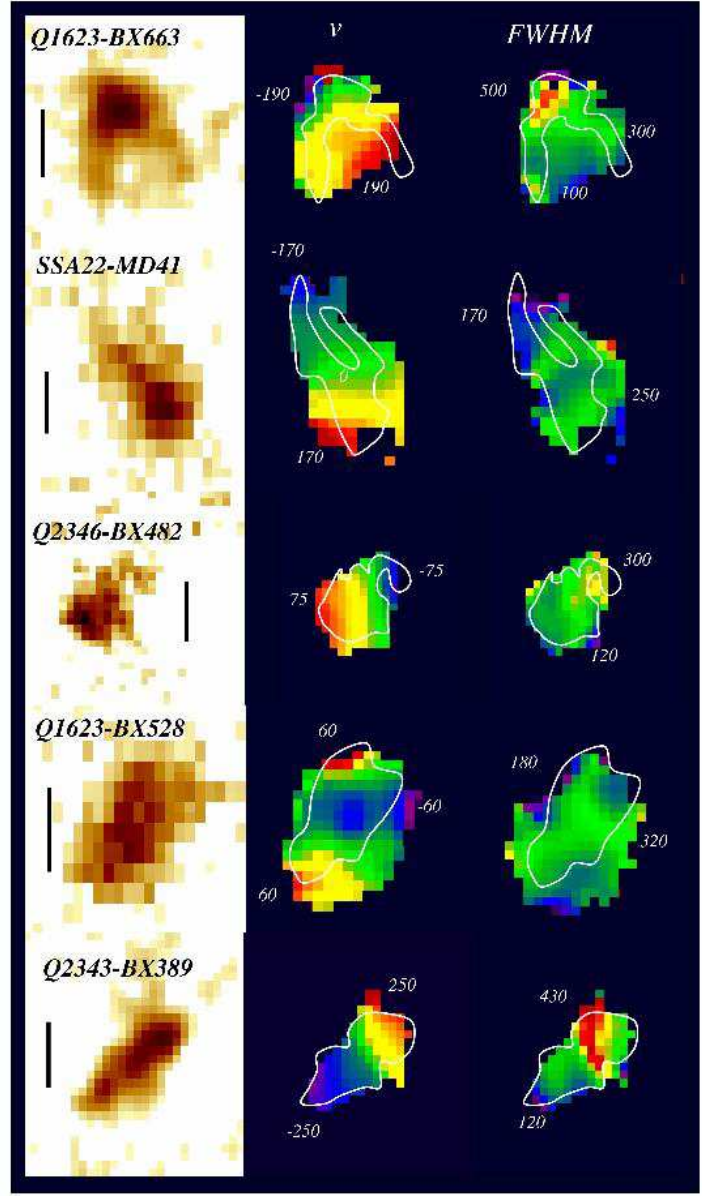}
\vspace{0.5cm}
\caption{
\small
Two-dimensional H$\alpha$ kinematics for five of our BM/BX sample galaxies
observed with SINFONI.  Each row corresponds to the results for one source.
The panels to the left show the integrated H$\alpha$ linemap, with the
angular scale of $1^{\prime\prime}$ indicated ({\em vertical bar\/}),
corresponding to $\rm \approx 8.3~kpc$ at the redshifts of the sources.
The middle panels show the spatial distribution of velocity $v$
(relative to the systemic velocity), obtained from Gaussian fits to
the data cubes after smoothing spatially with a two-dimensional Gaussian of
$\rm FWHM = 3~pixels$ (or 0\farcs 38, resulting in an effective resolution
of $\approx 0\farcs 6$).
The panels to the right show the spatial distribution of the FWHM velocity
width obtained from the same fits.
The velocity and FWHM maps are color-coded with a linear stretch and such
that values increase from magenta to red; minimum and maximum values for each
galaxy are labeled on the maps.  An outline of the integrated H$\alpha$
linemap is overplotted on the velocity and FWHM maps.
In all images, north is up and east is to the left.
\label{fig-maps2}
}
\end{figure}

\clearpage

\begin{figure}[p]
\figurenum{3}
\epsscale{0.72}
\plotone{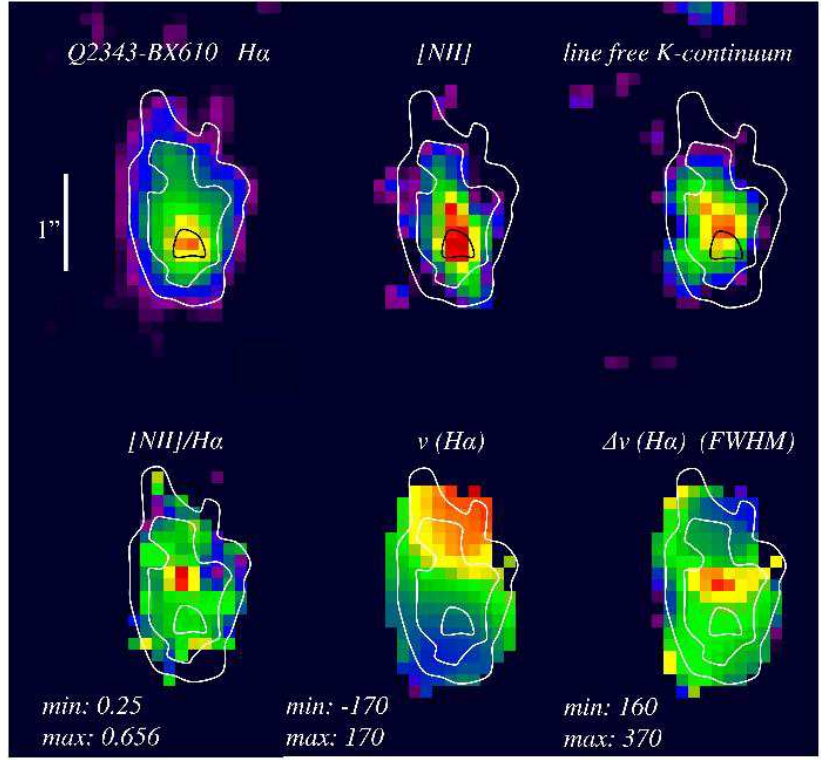}
\vspace{0.5cm}
\caption{
\small
Rest-frame optical emission and H$\alpha$ kinematics of $\rm Q2343-BX610$.
The maps shown are the integrated H$\alpha$ line emission ({\em top left\/}),
the integrated [\ion{N}{2}]\,$\lambda 6584$ line emission ({\em top center\/}),
the integrated line-free observed $K$ continuum emission ({\em top right\/}),
the [\ion{N}{2}]\,$\lambda 6584$/H$\alpha$ line ratio ({\em bottom left\/}),
the H$\alpha$ velocity field $v({\rm H\alpha})$ ({\em bottom center\/}),
and the H$\alpha$ FWHM velocity width $\Delta v({\rm H\alpha})$
({\em bottom right\/}).
The vertical bar left of the H$\alpha$ linemap is $1^{\prime\prime}$
in length, or 8.26~kpc at the redshift of $\rm Q2343-BX610$.
Each map is color-coded with a linear scaling and such that the
values increase from magenta to red.  The minimum and maximum values
of [\ion{N}{2}]\,$\lambda 6584$/H$\alpha$ ratio (unitless), and of
$v({\rm H\alpha})$ and $\Delta v({\rm H\alpha})$ (in $\rm km\,s^{-1}$)
are labeled below their respective maps.  All maps have been extracted
from the reduced data cube after smoothing spatially by a two-dimensional
Gaussian of $\rm FWHM = 3~pixels$ (or 0\farcs 38).
Contours of the H$\alpha$ line emission are overplotted on each map.
In all images, north is up and east is to the left.
\label{fig-bx610}
}
\end{figure}

\clearpage

\begin{figure}[p]
\figurenum{4}
\epsscale{0.95}
\plotone{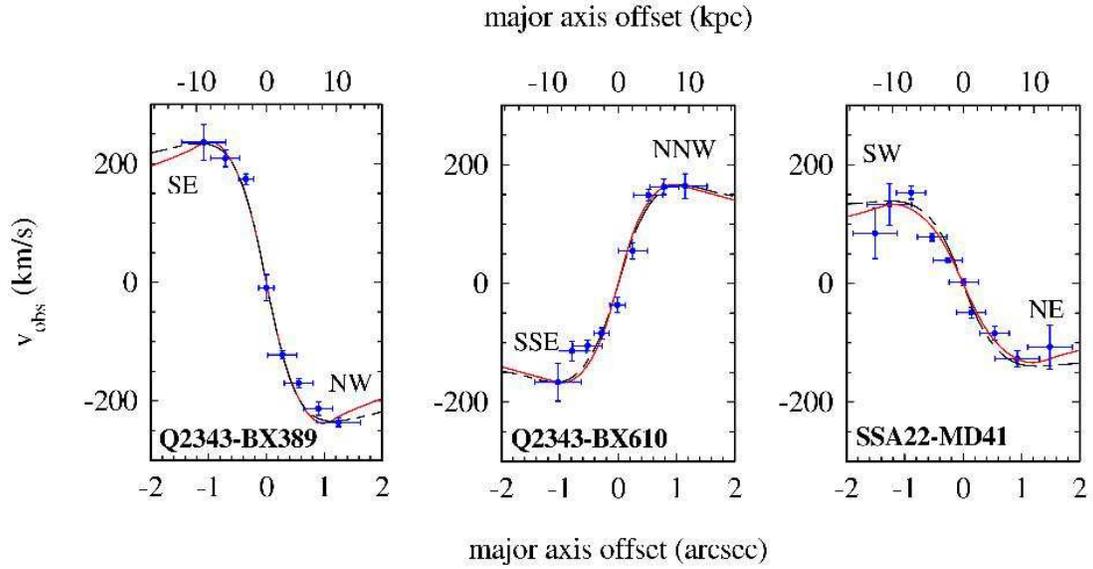}
\vspace{0.2cm}
\caption{
\small
Rotation curves of $\rm SSA22a-MD41$, $\rm Q2343-BX610$, and
$\rm Q2343-BX389$ derived from the H$\alpha$ line emission.
In each panel, the data points show the rotation velocities (relative to the
systemic velocity) as a function of position along the kinematic major axis
of each galaxy.  The velocities were obtained from Gaussian fits to the line
profiles measured in spectra extracted in synthetic circular apertures.
The horizontal error bars indicate the diameter of the synthetic apertures.
The vertical error bars correspond to the formal $1\sigma$ uncertainties of
the velocities.
The rotation curves of the disk models that best fit the observed H$\alpha$
kinematics of each galaxy and described in \S~\ref{Sect-case} are also plotted.
The best-fit ring models are plotted as red solid lines, and the best-fit
exponential disk models (with central hole) are shown as black dashed lines.
\label{fig-velprof}
}
\end{figure}


\begin{figure}[p]
\figurenum{5}
\epsscale{0.50}
\plotone{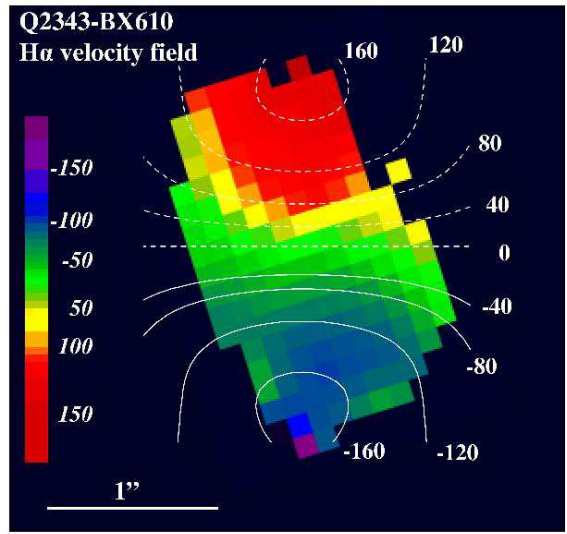}
\vspace{0.2cm}
\caption{
\small
Two-dimensional H$\alpha$ velocity field of $\rm Q2343-BX610$.
The velocity field derived from the observed H$\alpha$ line emission is
shown in colours, with a linear scaling increasing from magenta to red.
The superposed contours show the isovelocity map from the best-fit
rotating disk model with values relative to the systemic velocity
in $\rm km\,s^{-1}$ labeled (see \S~\ref{Sub-bx610}).
\label{fig-spider}
}
\end{figure}


\begin{thebibliography}{}

\bibitem[Abadi \etal(2003)]{Aba03} Abadi, M. G.,
         Navarro, J. F., Steinmetz, M., \& Eke, V. R.
         2003, \apj, 591, 499
\bibitem[Adelberger \etal(2005a)]{Ade05a} Adelberger, K. L., Erb, D. E.,
         Steidel, C. C., Reddy, N. A., Pettini, M., \& Shapley, A. E.
         2005a, \apj, 620, L75
\bibitem[Adelberger \etal(2005b)]{Ade05b} Adelberger, K. L., Steidel, C. C.,
         Pettini, M., Shapley, A. E., Reddy, N. A., \& Erb, D. E.
         2005b, \apj, 619, 697
\bibitem[Adelberger \etal(2004)]{Ade04} Adelberger, K. L., Steidel, C. C.,
         Shapley, A. E., Hunt, M. P., Erb, D. K., Reddy, N. A., \& Pettini, M.
         2004, \apj, 607, 226
\bibitem[Armus \etal(1990)Armus, Heckman, \& Miley]{Arm90}
         Armus, L., Heckman, T. M., \& Miley, G. K.
         1990, \apj, 364, 471
\bibitem[Asplund \etal(2004)]{Asp04} Asplund, M., Grevesse, N.,
         Sauval, A. J., Allende Prieto, C., \& Kiselman, D.
         2004, \aap, 417, 751
\bibitem[Baugh \etal(2005)]{Bau05} Baugh, C. M., Lacey, C. G., Frenk, C. S.,
         Granato, G. L., Silva, L., Bressan, A., Benson, A. J., \& Cole, S.
         2005, \mnras, 356, 1191
\bibitem[Bonnet \etal(2003)]{Bon03} Bonnet, H., \etal
         2003, Proc. SPIE, 4839, 329
\bibitem[Bonnet \etal(2004)]{Bon04} Bonnet, H., \etal
         2004, The Messenger, 117, 17
\bibitem[Cardone \& Sereno(2005)]{Car05} Cardone, V. F., \& Sereno, M.
         2005, \aap, 438, 545
\bibitem[Chabrier(2003)]{Cha03} Chabrier, G.
         2003, \pasp, 115, 763
\bibitem[Chapman \etal(2005)]{Cha05} Chapman, S. C.,
         Blain, A. W., Smail, I., \& Ivison, R. J.
         2005, \apj, 622, 772
\bibitem[Cowie \etal(1995) Cowie, Hu, \& Songaila]{Cow95}
         Cowie, L., Hu, E., \& Songaila, A.
         1995, \aj, 110, 1576
\bibitem[Dekel \& Birnboim(2006)]{Dek06} Dekel, A. \& Birnboim, Y.
         2006, \mnras, in press (astro-ph/0412300)
\bibitem[Dickinson \etal(2003)]{Dic03} Dickinson, M.,
         Papovich, C., Ferguson, H. C., \& Budav\'ari, T.
         2003, \apj, 587, 25
\bibitem[Downes \& Solomon(1998)]{Dow98} Downes, D., \& Solomon, P. M.
         1998, \apj, 507, 615
\bibitem[Eisenhauer \etal(2003a)]{Eis03a} Eisenhauer, F., \etal
         2003a, Proc. SPIE, 4841, 1548
\bibitem[Eisenhauer \etal(2003b)]{Eis03b} Eisenhauer, F., \etal
         2003b, The Messenger, 113, 17
\bibitem[Elmegreen(1999)]{Elm99} Elmegreen, B.G.,
         1999, \apj, 517, 103
\bibitem[Elmegreen \& Elmegreen(2005)]{Elm05}
         Elmegreen, B. G., Elmegreen, D. M.
         2005, \apj, 627, 632
\bibitem[Elmegreen \etal(2004a)Elmegreen, Elmegreen, \& Hirst]{Elm04a}
         Elmegreen, B. G., Elmegreen, D. M., \& Hirst, A. C.
         2004a, \apj, 612, 191
\bibitem[Elmegreen \etal(2004b)Elmegreen, Elmegreen, \& Sheets]{Elm04b}
         Elmegreen, D. M., Elmegreen, B. G., \& Sheets, C. M.
         2004b, \apj, 603, 74
\bibitem[Elmouttie \etal(1998)]{Elm98}
         Elmouttie, M., Koribalski, B., Gordon, S., Taylor, K.,
         Houghton, S., Lavezzi, T., Haynes, R., \& Jones, K.
         1998, \mnras, 297, 49
\bibitem[Erb \etal(2003)]{Erb03} Erb, D. K., Shapley, A. E.,
         Steidel, C. C., Pettini, M., Adelberger, K. L.,
         Hunt, M. P., Moorwood, A. F. M., \& Cuby, J.-G.
         2003, \apj, 591, 101
\bibitem[Erb \etal(2004)]{Erb04} Erb, D. K., Steidel, C. C.,
         Shapley, A. E., Pettini, M., \& Adelberger, K. L.
         2004, \apj, 612, 122
\bibitem[Erb \etal(2006a)]{Erb06a} Erb, D. K., Shapley, A. E.,
         Pettini, M., Steidel, C. C., Reddy, N. A., \& Adelberger, K. L.
         2006a, \apj, in press (astro-ph/0602473)
\bibitem[Erb \etal(2006b)]{Erb06b} Erb, D. K., Steidel, C. C.,
         Shapley, A. E., Pettini, M., Reddy, N. A., \& Adelberger, K. L.
         2006b, \apj, submitted
\bibitem[Fan \etal(2001)]{Fan01} Fan, X., \etal
         2001, \aj, 121, 54
\bibitem[F\"orster Schreiber \etal(2004)]{FS04}
         F\"orster Schreiber, N. M., \etal, 2004, \apj, 616, 40
\bibitem[Fontana \etal(2003)]{Fon03} Fontana, A., \etal
         2003, \apj, 594, L9
\bibitem[Franx \etal(2003)]{Fra03} Franx, M., \etal,
         2003, \apj, 587, L79
\bibitem[Fukunaga(1984)]{Fuk84} Fukunaga, M.
         1984, \pasp, 36, 433
\bibitem[Gammie \etal(1991)Gammie, Ostriker, \& Jog]{Gam91}
         Gammie, C. F., Ostriker, J. P., \& Jog, C. J.
         1991, \apj, 378, 565
\bibitem[Genzel \etal(2001)]{Gen01} Genzel, R.,
         Tacconi, L. J., Rigopoulou, D., Lutz, D., \& Tecza, M.
         2001, \apj, 563, 527
\bibitem[Gillessen \etal(2005)]{Gil05} Gillessen, S., \etal
         2005, The Messenger, 120, 26
\bibitem[Goldreich \& Tremaine(1978)]{Gol78} Goldreich, P., \& Tremaine, S.
         1978, Icarus, 34, 227
\bibitem[Gordon \etal(2000)]{Gor00}
         Gordon, S., Koribalski, B., Houghton, S., \& Jones, K.
         2000, \mnras, 315, 248
\bibitem[Hayashi \etal(2004)]{Hay04} Hayashi, E., \etal
         2004, \mnras, 355, 794
\bibitem[Heisler \etal(1985)Heisler, Tremaine, \& Bahcall]{Hei85}
         Heisler, J., Tremaine, S., \& Bahcall, J. N.
         1985, \apj, 298, 8
\bibitem[Immeli \etal(2004a)]{Imm04a}
         Immeli, A., Samland, M., Gerhard, O., \& Westera, P.
         2004a, \aap, 413, 547
\bibitem[Immeli \etal(2004b)]{Imm04b}
         Immeli, A., Samland, M., Westera, P., \& Gerhard, O.
         2004b, \apj, 611, 20
\bibitem[Jog \& Ostriker(1988)]{Jog88} Jog, C. J., \& Ostriker, J. P.
         1988, \apj, 328, 404
\bibitem[Kauffmann \etal(2003)]{Kau03} Kauffmann, G., \etal,
         2003, \mnras, 346, 1055
\bibitem[Kewley \etal(2001)]{Kew01} Kewley, L. J., Dopita, M. A.,
         Sutherland, R. S., Heisler, C. A., \& Trevena, J.
         2001, \apj, 556, 121
\bibitem[Kroupa(2001)]{Kro01} Kroupa, P.
         2001, \mnras, 322, 231
\bibitem[Labb\'e \etal(2005)]{Lab05} Labb\'e, I., \etal,
         2005, \apj, 624, L81
\bibitem[Law \etal(2006)Law, Steidel, \& Erb]{Law06}
         Law, D. R., Steidel, C. C., \& Erb, D. K.
         2006, \aj, 131, 70
\bibitem[Lehnert \& Heckman(1994)]{Leh94} Lehnert, M. D. \& Heckman, T. M.
         1994, \apj, 426, L27
\bibitem[Lehnert \& Heckman(1995)]{Leh95} Lehnert, M. D. \& Heckman, T. M.
         1995, \apjs, 97, 89
\bibitem[Lehnert \& Heckman(1996a)]{Leh96a} Lehnert, M. D. \& Heckman, T. M.
         1996a, \apj, 462, 651
\bibitem[Lehnert \& Heckman(1996b)]{Leh96b} Lehnert, M. D. \& Heckman, T. M.
         1996b, \apj, 472, 546
\bibitem[Leitherer \etal(1999)]{Lei99} Leitherer, C., \etal
         1999, \apjs, 123, 3
\bibitem[Mihos \& Bothun(1998)]{Mih98} Mihos, J. C., \& Bothun, G. D.
         1998, \apj, 500, 619
\bibitem[Mo \etal(1998)Mo, Mao, \& White]{Mo98}
         Mo, H. J., Mao, S. \& White, S. D. M.
         1998, \mnras, 295, 319
\bibitem[Mo \& White(2002)]{Mo02} Mo, H. J., \& White, S. D. M.
         2002, \mnras, 336, 112
\bibitem[Pagel \etal(1979)]{Pag79} Pagel, B. E. J.,
         Edmunds, M. G., Blackwell, D. E., Chun, M. S., \& Smith, G.
         1979, \mnras, 189, 95
\bibitem[Peebles(1969)]{Pee69} Peebles, P. J. E.
         1969, \apj, 155, 393
\bibitem[Pettini \& Pagel(2004)]{Pet04} Pettini, M., \& Pagel, B. E. J.
         2004, \mnras, 348, L59
\bibitem[Reddy \etal(2005)]{Red05} Reddy, N. A., Erb, D. K.,
         Steidel, C. C., Shapley, A. E., Adelberger, K. L., \& Pettini, M.
         2005, \apj, 633, 748
\bibitem[Rudnick \etal(2003)]{Rud03} Rudnick, G., \etal
         2003, \apj, 599, 847
\bibitem[Ryder(1995)]{Ryd95} Ryder, S.
         1995, \apj, 444, 610
\bibitem[Scalo(1986)]{Sca86} Scalo, J. M.
         1986, Fund. Cosmic Phys., 11, 1
\bibitem[Shapley \etal(2004)]{Sha04} Shapley, A. E., Erb, D. K.,
         Pettini, M., Steidel, C. C., \& Adelberger, K. L.
         2004, \apj, 612, 108
\bibitem[Shapley \etal(2005)]{Sha05} Shapley, A. E., Steidel, C. C.,
         Erb, D. K., Reddy, N. A., Adelberger, K. L., Pettini, M.,
         Barmby, P., \& Huang, J.
         2005, \apj, 626, 698
\bibitem[Spergel \etal(2003)]{Spe03} Spergel, D. N., \etal
         2003, ApJS, 148, 175
\bibitem[Steidel \etal(2005)]{Ste05} Steidel, C. C., Adelberger, K. L.,
         Shapley, A. E., Erb, D. K., Reddy, N. A., \& Pettini, M.
         2005, \apj, 626, 44
\bibitem[Steidel \etal(2004)]{Ste04} Steidel, C. C.,
         Shapley, A. E., Pettini, M., Adelberger, K. L.,
         Erb, D. K., Reddy, N. A., \& Hunt, M. P.
         2004, \apj, 604, 534
\bibitem[Tecza \etal(2000)]{Tec00} Tecza, M., Genzel, R., 
         Tacconi, L. J., Anders, S., Tacconi-Garman, L. E., \& Thatte, N.
         2000, \apj, 537, 178
\bibitem[Tecza \etal(2004)]{Tec04} Tecza, M., \etal
         2004, \apj, 605, L109
\bibitem[Thompson \etal(2005)Thompson, Quataert, \& Murray]{Tho05}
         Thompson, T. A., Quataert, E., \& Murray, N.
         2005, \apj, 630, 167
\bibitem[Toomre(1964)]{Too64} Toomre, A.
         1964, \apj, 139, 1217
\bibitem[Tran \etal(2003)]{Tra03} Tran, H. D., \etal
         2003, \apj, 585, 750
\bibitem[Tremonti \etal(2004)]{Tre04} Tremonti, C.A., Heckman, T.M.,
         Kauffmann, G., Brinchmann, J., Charlot, S., White, S.D.M.,
         Seibert, M., Peng, E., Schlegel, D.J., Uomoto, A., Fukugita,
         M. \& Brinkman, J. 2004, \apj, 613, 898
\bibitem[van der Kruit \& Allen(1978)]{vdK78}
         van der Kruit, P. C., \& Allen, R. J.
         1978, \araa, 16, 103
\bibitem[van Dokkum \etal(2003)]{Dok03} van Dokkum, P. G., \etal
         2003, \apj, 587, L83
\bibitem[van Dokkum \etal(2004)]{Dok04} van Dokkum, P. G., \etal
         2004, \apj, 611, 703
\bibitem[White(1984)]{Whi84} White, S. D. M.
         1984, \apj, 286, 38
\bibitem[White \& Frenk(1991)]{Whi91} White, S. D. M., \& Frenk, C. S.
         1991, \apj, 379, 52


\end{thebibliography}
\end{document}